\documentclass[preprint,11pt]{elsarticle}

\usepackage{amsthm} 
\usepackage{amssymb} 
\usepackage{amstext} 
\usepackage{amsmath} 
\usepackage{bm} 
\usepackage{graphicx} 
\usepackage{epstopdf} 
\usepackage{subcaption} 
\usepackage{float}
\usepackage{tikz}
\usepackage{lineno,hyperref}
\usepackage[capitalize]{cleveref} 
\usepackage[makeroom]{cancel} 
\usepackage{multirow} 
\usepackage{booktabs} 
\usepackage{lscape}
\usepackage{tabularx} 
\usepackage{adjustbox}
\usepackage{tablefootnote}
\usepackage{threeparttablex}
\usepackage{longtable}

\makeatletter
\newcommand{\mathleft}{\@fleqntrue\@mathmargin0pt}
\newcommand{\mathcenter}{\@fleqnfalse}
\makeatother

\modulolinenumbers[5]

\crefalias{subequation}{equation} 
\crefalias{eqnarray}{equation} 
\crefformat{pluraleq}{Eqs.~(#2#1#3)} 
\Crefformat{pluraleq}{Equations~(#2#1#3)} 

\def\bal#1\nal{\begin{align}#1\end{align}}
\def\bala#1\nala{\begin{align*}#1\end{align*}}
\def\bsub#1\nsub{\begin{subequations}#1\end{subequations}}
\newcommand{\f}{\frac}
\newcommand{\ux}{{\bm x}}

\newcommand{\uom}{{\bf \Omega}}

\journal{ }

 \biboptions{sort&compress}
\begin{document}

	\begin{frontmatter}
		
		\title{On the Application of the Analytical Discrete Ordinates Method to the Solution of Nonclassical Transport Problems in Slab Geometry}
		\author[uerj]{L.R.C. Moraes\corref{cor1}}
		\author[ufrgs]{L.B. Barichello\fnref{liliane}}
		\author[uerj]{R.C. Barros\fnref{ricardo}}
		\author[osu]{R. Vasques\fnref{richard}}
		\address[uerj]{Universidade do Estado do Rio de Janeiro, Departamento de Modelagem Computacional – IPRJ, Rua Bonfim 25, 28625-570, Nova Friburgo, RJ, Brazil}
		\address[ufrgs]{Universidade Federal do Rio Grande do Sul, Instituto de Matemática e Estatística,\\ Av. Bento Gonçalves 9500, 91509-900, Porto Alegre, RS, Brasil}
		\address[osu]{The Ohio State University, Department of Mechanical and Aerospace Engineering, 201 W. 19th Avenue, Columbus, OH 43210, United States of America}
		\cortext[cor1]{Corresponding author:  leonardrcmoraes@gmail.com}
		\fntext[liliane]{lbaric@mat.ufrgs.br}
		\fntext[ricardo]{ricardo.barros@uerj.br}
		\fntext[richard]{richard.vasques@fulbrightmail.org}
		\begin{abstract}
			In this work we investigate the use of the Analytical Discrete Ordinates (ADO) method when solving the spectral approximation of the nonclassical transport equation. The spectral approximation is a recently developed method based on the representation of the nonclassical angular flux as a series of Laguerre polynomials. This representation generates, as outcome, a system of equations that have the form of classical transport equations and can therefore be solved by current deterministic algorithms. Thus, the investigation of efficient approaches to solve the nonclassical transport equation is of interest and shall be pursued. This is the case of the ADO method which has been successfully used to solve a wide class of problems in the general area of particle transport. Numerical results are presented for two nonclassical test problems in slab geometry. These nonclassical transport problems are chosen in such way that their solution exactly reproduces the solution of the classical diffusion problem. Very accurate results are obtained for both test problems. However, the use of high precision arithmetic is sometimes required as illustrated in the second test problem. Limitations of the spectral approximation are also analyzed and discussed.

		\end{abstract}
		
	\end{frontmatter}

	\section{Introduction}\label{s1}
	
	The \emph{Nonclassical Theory} of linear particle transport was developed to model transport processes in which the particle flux is \emph{not} exponentially attenuated.
	In the nuclear engineering community, the interest for this type of nonclassical process originated in 2004 during a multidisciplinary conference in Computational Methods in Transport \cite{Graziani:2006:Computational}.
	At this conference, the mathematical similarities between radiative transfer through atmospheric clouds \cite{Davis:2006:Effective} and neutron transport in pebble-bed reactors (PBRs)~\cite{Wu:2002:TheDesign,Koster:2003:PBMR,Kadak:2007:MIT} became apparent; specifically, the fact that correlations between scatterers and/or unresolved spatial fluctuations in the system lead to nonexponential decay of the particle flux.
	Since classical linear transport models inherently assume an exponential attenuation in the system, a nonclassical theory capable of addressing these issues needed to be derived, which prompted a generalization of the linear Boltzmann equation \cite{Larsen:2007:Generalized,Larsen:2011:Generalized}.
	
	Let $\boldsymbol{x} = (x,y,z)$ describe the location of a particle in space and $\boldsymbol{\Omega} = (\Omega_{x},\Omega_{y},\Omega_{z})$ represent the particle's direction of flight, with $|\boldsymbol{\Omega}|=1$.
	We define $s$ as the distance traveled by the particle since its last interaction (birth or scattering), such that $s=0$ at the interaction point.   
	The steady state, one-speed nonclassical linear Boltzmann equation with angular-dependent free-paths, as derived in \cite{Vasques:2014:NonclassicalI}, is given by
	\begin{subequations}\label[pluraleq]{eq1.1}
		\begin{align}
			\frac{\partial}{\partial s}&\Psi(\boldsymbol{x},\boldsymbol{\Omega},s) + \boldsymbol{\Omega}\boldsymbol{\cdot}\nabla\Psi(\boldsymbol{x},\boldsymbol{\Omega},s)+ \Sigma_{t}(\boldsymbol{\Omega},s)\Psi(\boldsymbol{x},\boldsymbol{\Omega},s) = \label{eq1.1a} 
			\\&
			\delta(s)\left[c\int_{4\pi}^{}\int_{0}^{\infty}P(\boldsymbol{\Omega}'\boldsymbol{\cdot}\boldsymbol{\Omega})\Sigma_{t}(\boldsymbol{\Omega}',s')\Psi(\boldsymbol{x},\boldsymbol{\Omega}',s')ds'd\Omega' + \frac{Q(\boldsymbol{x})}{4\pi}  \right], \quad \boldsymbol{x} \in V, \:\boldsymbol{\Omega}\in 4\pi, \: 0<s, \nonumber
		\end{align}
		where $\Psi$ is the nonclassical angular flux; $Q$ is an isotropic source; $\Sigma_t$ is the macroscopic total cross section; $c$ is the scattering ratio; and $P(\boldsymbol{\Omega}'\boldsymbol{\cdot}\boldsymbol{\Omega})d\Omega$ represents the probability that when a particle traveling with direction $\uom'$ scatters, its outgoing direction of flight will lie in a differential surface $d\Omega$ about $\uom$.
		
		The appropriate way to define boundary conditions to \cref{eq1.1a} is not universally agreed upon, due to the boundaries not being correlated with the positions of the scatterers (detailed discussion can be found in \cite{Frank:2015:Nonclassical}). 
		Nevertheless, a compelling case has been made \cite{Larsen:2017:Equivalence} to use 
		\begin{align}\label{eq1.1b}
			&\Psi(\boldsymbol{x},\boldsymbol{\Omega},s) = \Psi^{b}(\boldsymbol{x},\boldsymbol{\Omega})\delta(s), \quad \boldsymbol{x} \in \partial V,\: \boldsymbol{n}\boldsymbol{\cdot}\boldsymbol{\Omega}<0,\: 0<s.
		\end{align}
	\end{subequations}
	The Dirac delta function $\delta(s)$ in \cref{eq1.1} is used to ``reset" the value of $s$ for particles that have just scattered or been born in the system through $Q$ or $\Psi^{b}$, since at that moment their distance from the event is $s=0$.
	
	We remark that the macroscopic total cross section $\Sigma_t$ in the nonclassical formulation is a function of both $s$ and $\uom$.
	Specifically, $\Sigma_t(\uom,s)ds$ describes the probability that a particle, born or scattered at any position $\ux$ and with direction of flight given by $\uom$, will experience its next collision between $\ux+s\uom$ and $\ux+(s+ds)\uom$.
	It satisfies \cite{Vasques:2014:NonclassicalI}
	\begin{align}\label{eq1.2}
		p(\boldsymbol{\Omega},s) = \Sigma_{t}(\boldsymbol{\Omega},s)e^{-\int_{0}^{s}\Sigma_{t}(\boldsymbol{\Omega},s')ds'},
	\end{align}
	where $p(\boldsymbol{\Omega},s)$ is the free-path conditional distribution function in a given direction $\boldsymbol{\Omega}$.
	
	If the macroscopic total cross section is independent of $s$ and $\uom$, then \textit{classical} transport takes place, and \cref{eq1.2} reduces to the exponential distribution.
	In this case, \cref{eq1.1} reduce to the classical steady state, one-speed linear Boltzmann equations
	\begin{subequations}\label[pluraleq]{eq1.3}
		\begin{align}
			\boldsymbol{\Omega}\boldsymbol{\cdot}\nabla\tilde{\Psi}(\boldsymbol{x},\boldsymbol{\Omega})+ \sigma_{t}\tilde{\Psi}(\boldsymbol{x},\boldsymbol{\Omega}) &= c\int_{4\pi}^{}P(\boldsymbol{\Omega}'\boldsymbol{\cdot}\boldsymbol{\Omega})\sigma_{t}\tilde{\Psi}(\boldsymbol{x},\boldsymbol{\Omega}')d\Omega' + \frac{Q(\boldsymbol{x})}{4\pi}  , \quad \boldsymbol{x} \in V, \:\boldsymbol{\Omega}\in 4\pi, \label{eq1.3a}
			\\[5pt]
			\tilde{\Psi}(\boldsymbol{x},\boldsymbol{\Omega}) &= \tilde{\Psi}^{b}(\boldsymbol{x},\boldsymbol{\Omega}), \quad \boldsymbol{x} \in \partial V,\: \boldsymbol{n}\boldsymbol{\cdot}\boldsymbol{\Omega}<0,
		\end{align}
		where $\tilde{\Psi}$ is the classical angular flux given by
		\begin{align}\label{eq1.3c}
			\tilde{\Psi}(\boldsymbol{x},\boldsymbol{\Omega}) = \int_{0}^{\infty}\Psi(\boldsymbol{x},\boldsymbol{\Omega},s)ds.
		\end{align}
	\end{subequations}
	In \cref{eq1.3a} we represent the macroscopic total cross section by the variable $\sigma_{t}$ instead of $\Sigma_{t}$. This distinction is made to emphasize the classical meaning of the macroscopic total cross section when is independent of both $\boldsymbol{\Omega}$ and $s$. Moving forward, we use these two notations, $\Sigma_{t}$ and $\sigma_{t}$, to represent the macroscopic total cross section in the nonclassical and classical senses, respectively.
	
	Recently, a spectral method has been developed \cite{Vasques:2020:Spectral} to represent the nonclassical angular flux as a series of Laguerre polynomials in $s$.
	This method produces a system of equations that have the form of classical transport equations and can therefore be solved by current deterministic algorithms.
	In short, we define $\psi$ such that
	\begin{subequations}\label[pluraleq]{eq1.4}
		\begin{align}\label{eq1.4a}
			\Psi(\boldsymbol{x},\boldsymbol{\Omega},s) \equiv \psi(\boldsymbol{x},\boldsymbol{\Omega},s)e^{-\int_{0}^{s}\Sigma_{t}(\boldsymbol{\Omega},s')ds'},
		\end{align}
		and expand it as a series of Laguerre polynomials in $s$
		\begin{align}\label{eq1.4b}
			\psi(\boldsymbol{x},\boldsymbol{\Omega},s) = \sum_{m=0}^{\infty}\psi_{m}(\boldsymbol{x},\boldsymbol{\Omega})L_{m}(s),
		\end{align}
		where $L_{m}(s)$ is the Laguerre polynomial of order $m$.   
	\end{subequations}
	As shown in \cite{Vasques:2020:Spectral}, we can use this along with \cref{eq1.1} to obtain a system of equations for $\psi_m$
	\begin{subequations}\label[pluraleq]{eq1.5}
		\begin{align}
			\boldsymbol{\Omega}\boldsymbol{\cdot}\nabla\psi_{m}(\boldsymbol{x},\boldsymbol{\Omega}) + \sum_{j=0}^{m}\psi_{j}(\boldsymbol{x},\boldsymbol{\Omega}) &= c\int_{4\pi}^{}P(\boldsymbol{\Omega}'\boldsymbol{\cdot}\boldsymbol{\Omega})\sum_{k=0}^{\infty}\psi_{k}(\boldsymbol{x},\boldsymbol{\Omega}')\mathcal{L}_{k}(\boldsymbol{\Omega}')d\Omega' + \frac{Q(\boldsymbol{x})}{4\pi},\: \label{eq1.5a}
			\\[5pt] 
			\psi_{m}(\boldsymbol{x},\boldsymbol{\Omega}) &= \Psi^{b}(\boldsymbol{x},\boldsymbol{\Omega}),\quad \boldsymbol{x} \in \partial V,\: \boldsymbol{n}\boldsymbol{\cdot}\boldsymbol{\Omega}<0,
		\end{align}
		with $m=0,1,2,\dots,$ and
		\begin{align}\label{eq1.5c}
			\mathcal{L}_{k}(\boldsymbol{\Omega}) = \int_{0}^{\infty}p(\boldsymbol{\Omega},s')L_{k}(s')ds'.
		\end{align}
	\end{subequations}
	Once the series expansion is truncated, these equations can be solved through traditional deterministic approaches.
	The classical angular flux can be recovered using \cref{eq1.3c,eq1.4}.
	
	Assuming isotropic scattering and an angular-independent free-path distribution, we can write $P(\uom'\cdot\uom) = 1/4\pi$, $p(\uom,s) = p(s)$, and $\Sigma_t(\uom,s) = \Sigma_t(s)$.
	In this case, \cref{eq1.1a} simplifies to
	\begin{align} \label{eq1.6}
		\frac{\partial}{\partial s}&\Psi(\boldsymbol{x},\boldsymbol{\Omega},s) + \boldsymbol{\Omega}\boldsymbol{\cdot}\nabla\Psi(\boldsymbol{x},\boldsymbol{\Omega},s)+ \Sigma_{t}(s)\Psi(\boldsymbol{x},\boldsymbol{\Omega},s) =\\[5pt]
		& \frac{\delta(s)}{4\pi}\left[c\int_{4\pi}^{}\int_{0}^{\infty}\Sigma_{t}(s')\Psi(\boldsymbol{x},\boldsymbol{\Omega}',s')ds'd\Omega' + Q(\boldsymbol{x})  \right], \quad \boldsymbol{x} \in V, \:\boldsymbol{\Omega}\in 4\pi, \: 0<s\,.\nonumber
	\end{align}
	It has been shown \cite{Frank:2015:Nonclassical,Vasques:2016:Nonclassical,Makine:2018:Exact} that certain diffusion-based approximations to the classical and nonclassical linear Boltzmann equations can be represented exactly by \cref{eq1.6} when $\Sigma_t(s)$ is appropriately chosen.
	If we define the collision-rate density $f(\ux)$ such that
	\begin{align}\label{eq1.7}
		f(\boldsymbol{x}) = \int_{4\pi}\int_{0}^{\infty}\Sigma_{t}(s')\Psi(\boldsymbol{x},\boldsymbol{\Omega}',s')ds'd\Omega\,,
	\end{align}
	then \cref{eq1.6} can be manipulated into the following integral equation \citep{Frank:2015:Nonclassical}:
	\begin{align}\label{eq1.8}
		f(\boldsymbol{x}) = \int\int\int\left[cf(\boldsymbol{x}')+Q(\ux')\right]\frac{p(|\boldsymbol{x}'-\boldsymbol{x}|)}{4\pi|\boldsymbol{x}'-\boldsymbol{x}|^{2}}dV'\,.
	\end{align}
	Here, $p(|\boldsymbol{x}'-\boldsymbol{x}|)$ is the free-path distribution function.
	We can derive a similar expression for diffusion.
	The one-speed, classical diffusion equation with isotropic scattering is given by
	\begin{align}\label{eq1.9}
		-\frac{1}{3\sigma_{t}}\nabla^{2}\Phi(\boldsymbol{x})+\sigma_{t}\Phi(\boldsymbol{x}) = c\sigma_{t}\Phi(\boldsymbol{x}) + Q(\boldsymbol{x})\,,
	\end{align}
	where $\Phi$ is the classical scalar flux
	\begin{align}
		\Phi(\boldsymbol{x}) = \int_{4\pi}^{}\Psi(\boldsymbol{x},\boldsymbol{\Omega})d\Omega\,.
	\end{align} 
	Applying Green's function analysis to \cref{eq1.9}, one can obtain the collision-rate density \citep{Frank:2015:Nonclassical}
	\begin{align}\label{eq1.11}
		\sigma_{t}\Phi(\boldsymbol{x}) = f(\ux) =  \int\int\int\left[cf(\ux') + Q(\ux')\right]\frac{3\sigma^{2}_{t}|\boldsymbol{x}-\boldsymbol{x}'|e^{-\sqrt{3}\sigma_{t}|\boldsymbol{x}-\boldsymbol{x}'|}}{4\pi|\boldsymbol{x}-\boldsymbol{x}'|^{2}}dV'.
	\end{align}
	Comparing \cref{eq1.8} with \cref{eq1.11}, we see that they are the same if and only if
	\begin{subequations}
		\begin{align}\label{eq1.12a}
			p(s) = \lambda^{2}se^{-\lambda s},
		\end{align}
		where
		\begin{align}\label{eq1.12b}
			\lambda = \sqrt{3}\sigma_{t}. 
		\end{align}
		In this case, the nonclassical function $\Sigma_{t}(s)$ is given by
		\begin{align}\label{eq1.12c}
			\Sigma_{t}(s) = \frac{\lambda^{2}s}{1+\lambda s}.
		\end{align}
	\end{subequations}
	
	In this work we present a detailed study of the spectral approach for solving the nonclassical transport equation applied to diffusion.
	We have opted to focus on the classical diffusion problem for three reasons: (i) its theory has been  greatly explored and well-documented by several authors \cite{Bell:1970:Nuclear,Shqair:2019:Analytical,Nahla:2012:Advanced,Lee:2020:Numerical}, (ii) its solution can be represented exactly by solving the appropriate nonclassical transport equation \cite{Frank:2010:Generalized,Vasques:2016:Nonclassical}, as detailed in the previous discussion; and (iii) the functions $\mathcal{L}_{k}$ can be calculated analytically considering the free-path distribution function as given in \cref{eq1.12a}.
	
	There are two main original contributions in this paper. The first one is a convergence analysis of the $\mathcal{L}_k$ functions when $p(s)$ is given by \cref{eq1.12a}, which throws new light on the numerical limitations of the spectral approach. The second is an investigation of the use of the Analytical Discrete Ordinates (ADO) method \cite{Barichello:1999:Discrete} to solve the slab geometry representation of the nonclassical problem described by \cref{eq1.5}. As the spectral method produces a system of equations that are suitable to the use of current deterministic methods, the exploration of efficient approaches to solve this problem is of interest and this work is a first step in this direction. In reference \cite{Vasques:2020:Spectral}, the classical Diamond Difference method \cite{LeMi93} was used along with this spectral decomposition, and numerical challenges pointed out the need of further investigation. To our knowledge, this is the first time the ADO method is applied to obtain solutions for the nonclassical transport equations.
	
	The ADO method has been successfully used to solve a wide class of problems in the general area of particle transport \cite{Barichello:1999:Discrete,Barichello:2011:Explicit,Barichello:2000:Particular,Barichelo:2019:Ontheuse,Barichello2000Jan,Barichello2001May}.
	Its main features include the generation of an explicit solution in the spatial variable and the use of arbitrary angular quadrature schemes, defined in the half-range interval, which determine an eigenvalue problem whose order is half the number of discrete angles.
	
	The remainder of this paper is organized as follows.
	In \cref{s2}, we present the convergence analysis of the $\mathcal{L}_k$ functions for the classical diffusion problem.
	In \cref{s3}, we give a detailed description of the ADO method as it is used to solve the nonclassical transport problem in slab geometry.
	\Cref{s4} introduces two test problems and presents numerical results.
	We discuss and analyze the accuracy and precision  of these results, describing some challenges that may arise from the use of the spectral approximation, and the application of the ADO method as well.
	Finally, in \cref{s5}, we conclude the paper with a brief discussion of the results and the prospects of future work.

	\section{Convergence of the $\mathcal{L}_{k}$ functions for the diffusion problem}\label{s2}
	\setcounter{equation}{0}
	
	As mentioned in the previous Section, classical diffusion modeled by \cref{eq1.9} can be represented exactly by \cref{eq1.6} if $\Sigma_t(s)$ is given by \cref{eq1.12c}.
	(This result is discussed in greater detail in \citep{Frank:2015:Nonclassical}).
	Therefore, using \cref{eq1.7,eq1.11}) we obtain
	\begin{align}\label{eq2.1}
		\Phi(\boldsymbol{x}) = \frac{1}{\sigma_{t}}\int_{4\pi}\int_{0}^{\infty}\Sigma_{t}(s')\Psi(\boldsymbol{x},\boldsymbol{\Omega}',s')ds'd\Omega',
	\end{align}
	which allows us to calculate the scalar flux solution of the classical diffusion equation using the solution of the nonclassical transport equation.  
	
	Using the spectral approach discussed in \cref{eq1.4}, we can rewrite \cref{eq2.1} as
	\begin{subequations}\label[pluraleq]{eq2.2}
		\begin{align}\label{eq2.2a}
			\Phi(\boldsymbol{x}) = \frac{1}{\sigma_{t}}\sum_{k=0}^{\infty}\mathcal{L}_{k}\int_{4\pi}\psi_{k}(\boldsymbol{x},\boldsymbol{\Omega})d\Omega\,.
		\end{align}
		Here, $\psi_{k}(\boldsymbol{x},\boldsymbol{\Omega})$ is obtained by solving \cref{eq1.5} while considering $p(s)$ as given by \cref{eq1.12a}, such that
		\begin{align}\label{eq2.2b}
			\mathcal{L}_{k} = \int_{0}^{\infty}\lambda^{2}se^{-\lambda s}L_{k}(s)ds\,.
		\end{align}
	\end{subequations}
	In summary, it is necessary to solve the improper integral on the right-hand side of this equation in order to obtain the scalar flux.
	
	To analyze the convergence of the $\mathcal{L}_{k}$ functions, we begin with representing the Laguerre polynomials $L_k(s)$ as \cite{Hochstrasser:1964:Orthogonal}
	\begin{align}\label{eq2.3}
		L_{k}(s) = \sum_{i=0}^{k} (-1)^{i}\binom{k}{i} \frac{s^{i}}{i!}\,.
	\end{align} 
	Substituting \cref{eq2.3} into \cref{eq2.2b}, we obtain
	\begin{align}\label{eq2.4}
		\mathcal{L}_{k} = \lambda^{2}\sum_{i=0}^{k}(-1)^{i}
		\binom{k}{i}\frac{1}{i!}\int_{0}^{\infty}s^{i+1}e^{-\lambda s}ds\,.
	\end{align}
	Using the change of variables $t=\lambda s$ and integrating by parts, we find that the improper integral above yields \cite{Davis:1964:Gamma}
	\begin{align}\label{eq2.5}
		\int_{0}^{\infty}s^{i+1}e^{-\lambda s}ds = \frac{(i+1)!}{\lambda^{i+2}}\,.
	\end{align}
	Substituting this result into \cref{eq2.4}, we obtain
	\begin{subequations}
		\begin{align}\label{eq2.6a}
			\mathcal{L}_{k} = \sum_{i=0}^{k}\tau^{i}\binom{k}{i}(i+1),
		\end{align} 
		where
		\begin{align}\label{eq2.6b}
			\tau = -1/\lambda.
		\end{align}
	\end{subequations}
	
	Now we rewrite \cref{eq2.6a} as the sum of two terms, $I_{1}$ and $I_{2}$, such that
	\begin{align}\label{eq2.7}
		\mathcal{L}_{k} = \underbrace{\sum_{i=0}^{k}\tau^{i}\binom{k}{i}}_{ I_{1}} + \underbrace{\sum_{i=0}^{k}i\tau^{i}\binom{k}{i}}_{I_{2}}\,.
	\end{align}
	Using the binomial theorem \cite{Goldberg:1964:Combinatorial}, we see that
	\begin{align}\label{eq2.8}
		I_{1} = \left(1+\tau\right)^{k}.
	\end{align}
	For the second term, we define $i=n+1$ and write
	\begin{align}\label{eq2.9}
		I_{2} = \sum_{n=0}^{k-1}(n+1)\binom{k}{n+1}\tau^{n+1},
	\end{align} 
	since the term $i=0$ (or $n=-1$) is 0.
	Using the binomial property
	\begin{align*}
		\binom{k}{n+1} = \binom{k}{n}\frac{k-n}{n+1}\,
	\end{align*}
	which is 0 when $n=k$, \cref{eq2.9} appear as
	\begin{align}\label{eq2.10}
		I_{2} = \tau\sum_{n=0}^{k} (k-n)\binom{k}{n}\tau^{n}\,.
	\end{align}	
	From the binomial theorem, we have
	\begin{align}\label{eq2.11}
		I_{2} = \tau k\left(1+\tau\right)^{k} - \tau\underbrace{\left[\sum_{n=0}^{k}n\tau^{n}\binom{k}{n}\right]}_{I_{2}}\,,
	\end{align}
	and hence
	\begin{align}\label{eq2.12}
		I_{2} = \tau k(1+\tau)^{k-1}.
	\end{align}	
	\Cref{eq2.8,eq2.12} allow us to rewrite \cref{eq2.7} as
	\begin{align}
		\mathcal{L}_{k} = \left(1+\tau \right)^{k} + \tau k\left(1+\tau \right)^{k-1}.
	\end{align}
	Finally, using \cref{eq1.12b,eq2.6b}, we obtain
	\begin{align}\label{eq2.14}
		\mathcal{L}_{k} = \left( 1-\frac{1}{\sqrt{3}\sigma_{t}}\right)^{k} - \frac{k}{\sqrt{3}\sigma_{t}}\left( 1-\frac{1}{\sqrt{3}\sigma_{t}}\right)^{k-1}.
	\end{align}
	
	We see that as $k \rightarrow \infty$, the functions $\mathcal{L}_{k}$ converge (to zero) only if $\sigma_{t} > \frac{\sqrt{3}}{6}$. 
	This introduces a limitation in the numerical procedure for problems in which $\sigma_{t} \leq \frac{\sqrt{3}}{6}$, since the $\mathcal{L}_{k}$ functions will diverge and the solution will not be attainable.
	This is further discussed within the context of the specific test problems in \cref{s42}.
	
	\section{An analytical discrete ordinates solution}\label{s3}
	\setcounter{equation}{0}
	In this section, we discuss the application of the ADO method to solve \cref{eq1.5} for problems in slab geometry, with  isotropic scattering and vacuum boundary conditions.
	Under these assumptions, and taking $M$ as the truncation order for the Laguerre series, we write \cref{eq1.5} as
	\begin{subequations}\label[pluraleq]{eq3.1}
		\begin{align}\label{eq3.1a}
			\frac{\partial}{\partial x} \psi_{m}(x,\mu) + \sum_{j=0}^{m}\psi_{j}(x,\mu) &= \frac{c}{2}\sum_{k=0}^{M}\int_{-1}^{1}\psi_{k}(x,\mu')\mathcal{L}_{k}(\mu')d\mu' + \frac{Q(x)}{2},\\
			\label{eq3.1b}
			\psi_{m}(0,\mu) &= 0,\quad \mu > 0\,,\\
			\label{eq3.1c}
			\psi_{m}(X,\mu) &= 0,\quad \mu < 0\:,
		\end{align}
	\end{subequations}
	where $m = 0,1,\dots,M$. As the problem stated by \cref{eq3.1} is linear, we write its general solution as a superposition of the homogeneous and particular solutions of \cref{eq3.1a} \cite{Barichello:1999:Discrete,Barichello:2011:Explicit,Pazinatto:2016:Analytical,Neto:2018:Problema,Barichello2000Jan,Barichello2001May}.
	
	\subsection{Homogeneous solution}\label{s3.1}
	To begin, we write the homogeneous version of \cref{eq3.1a} in a convenient matrix form as
	\begin{subequations}\label[pluraleq]{eq3.2}
		\begin{align}\label{eq3.2a}
			\mu\frac{\partial}{\partial x}\boldsymbol{\psi}^{h}(x,\mu) + \boldsymbol{\mathcal{T}}\boldsymbol{\psi}^{h}(x,\mu) = \frac{c}{2}\int_{0}^{1}\left[\boldsymbol{L}(\mu')\boldsymbol{\psi}^{h}(x,\mu') + \boldsymbol{L}(-\mu')\boldsymbol{\psi}^{h}(x,-\mu')\right]d\mu', 
		\end{align}
		where $\boldsymbol{\psi}^{h}(x,\mu)$ is a $M$-dimensional vector composed of the homogeneous solutions, such that,
		\begin{align}\label{eq3.2b}
			\boldsymbol{\psi}^{h}(x,\mu) = \left[\psi^{h}_{0}(x,\mu),\:\psi^{h}_{1}(x,\mu),\:\dots,\:\psi^{h}_{M}(x,\mu) \right]^{T}, 
		\end{align}
		$\boldsymbol{\mathcal{T}}$ is a lower triangular matrix of order $M$ whose non-zero entries are equal to one and $\boldsymbol{L}(\mu)$ is a square matrix of order $M$ defined as 
		\begin{align}\label{eq3.2c}
			\boldsymbol{L}(\mu) = \left[\boldsymbol{\mathcal{L}}_{0}(\mu),\: \boldsymbol{\mathcal{L}}_{1}(\mu),\: \dots,\: \boldsymbol{\mathcal{L}}_{M}(\mu) \right],
		\end{align}
		with $\boldsymbol{\mathcal{L}}_{m}(\mu)$ being an $M$-dimensional vector whose entries are $\mathcal{L}_{m}(\mu)$.
	\end{subequations}
	The superscript $T$ in \cref{eq3.2b} is used in this work to indicate the vector transpose. Furthermore, following the ADO procedure \cite{Barichello:1999:Discrete} we consider a quadrature scheme defined in the semi-interval $[0, 1]$, formed by $N$ nodes $\mu_{n}$ and corresponding weights $\omega_{n}$, to write \cref{eq3.2a} as a linear system composed of $2MN = 2\times M \times N$ ordinary differential equations. That is,
	\begin{subequations}\label[pluraleq]{eq3.3}	
		\begin{align}\label{eq3.3a}
			\mu_{n}\frac{d}{dx}\boldsymbol{\psi}^{h}(x,\mu_{n}) + \boldsymbol{\mathcal{T}}\boldsymbol{\psi}^{h}(x,\mu_{n}) = \frac{c}{2}\sum_{i=1}^{N}\left[\boldsymbol{L}(\mu_{i})\boldsymbol{\psi}^{h}(x,\mu_{i}) + \boldsymbol{L}(-\mu_{i})\boldsymbol{\psi}^{h}(x,-\mu_{i})\right]\omega_{i}
		\end{align}
		and
		\begin{align}\label{eq3.3b}
			-\mu_{n}\frac{d}{dx}\boldsymbol{\psi}^{h}(x,-\mu_{n}) + \boldsymbol{\mathcal{T}}\boldsymbol{\psi}^{h}(x,-\mu_{n}) = \frac{c}{2}\sum_{i=1}^{N}\left[\boldsymbol{L}(\mu_{i})\boldsymbol{\psi}^{h}(x,\mu_{i}) + \boldsymbol{L}(-\mu_{i})\boldsymbol{\psi}^{h}(x,-\mu_{i})\right]\omega_{i},
		\end{align}
	\end{subequations}
	where $n = 1,2,\dots,N$.
	
	Following the literature \cite{Barichello:1999:Discrete,Barichello:2011:Explicit}, we seek homogeneous solutions of the form
	\begin{align}\label{eq3.4}
		\boldsymbol{\psi}^{h}(x,\mu) = \boldsymbol{\phi}(\vartheta,\mu)e^{-\frac{x}{\vartheta}},
	\end{align}
	where $\vartheta$ is a constant and $\boldsymbol{\phi}(\vartheta,\mu)$ is an $M$-dimensional vector defined as
	\begin{align}\label{eq3.5}
		\boldsymbol{\phi}(\vartheta,\mu) = \left[\phi_{0}(\vartheta,\mu),\: \phi_{1}(\vartheta,\mu),\: \dots,\: \phi_{M}(\vartheta,\mu) \right]^{T}.
	\end{align}
	Substituting \cref{eq3.4} into \cref{eq3.3} we obtain
	\begin{subequations}\label[pluraleq]{eq3.6}
		\begin{align}\label{eq3.6a}
			\left( \boldsymbol{\mathcal{T}} - \frac{\mu_{n}}{\vartheta}\boldsymbol{I}_{M}\right)\boldsymbol{\phi}(\vartheta,\mu_{n}) = \frac{c}{2}\sum_{i=0}^{N}\left[\boldsymbol{L}(\mu_{i})\boldsymbol{\phi}(\vartheta,\mu_{i}) + \boldsymbol{L}(-\mu_{i})\boldsymbol{\phi}(\vartheta,-\mu_{i})\right]\omega_{i}
		\end{align}
		and
		\begin{align}\label{eq3.6b}
			\left( \boldsymbol{\mathcal{T}} + \frac{\mu_{n}}{\vartheta}\boldsymbol{I}_{M}\right)\boldsymbol{\phi}(\vartheta,-\mu_{n}) = \frac{c}{2}\sum_{i=0}^{N}\left[\boldsymbol{L}(\mu_{i})\boldsymbol{\phi}(\vartheta,\mu_{i}) + \boldsymbol{L}(-\mu_{i})\boldsymbol{\phi}(\vartheta,-\mu_{i})\right]\omega_{i},
		\end{align}
	\end{subequations}
	where $\boldsymbol{I}_{M}$ is the identity matrix of order $M$. By varying $n$ from $1$ to $N$ in \cref{eq3.6} we obtain
	\begin{subequations}\label[pluraleq]{eq3.7}
		\begin{align}\label{eq3.7a}
			\left(\boldsymbol{D} - \frac{1}{\vartheta}\boldsymbol{M} \right)\boldsymbol{\Phi}_{+}(\vartheta) = \frac{c}{2}\left[\boldsymbol{K}_{+}\boldsymbol{\Phi}_{+}(\vartheta) + \boldsymbol{K}_{-}\boldsymbol{\Phi}_{-}(\vartheta) \right] 
		\end{align}
		and 
		\begin{align}\label{eq3.7b}
			\left(\boldsymbol{D} + \frac{1}{\vartheta}\boldsymbol{M} \right)\boldsymbol{\Phi}_{-}(\vartheta) = \frac{c}{2}\left[\boldsymbol{K}_{+}\boldsymbol{\Phi}_{+}(\vartheta) + \boldsymbol{K}_{-}\boldsymbol{\Phi}_{-}(\vartheta) \right].
		\end{align}
		In \cref{eq3.7a,eq3.7b} $\boldsymbol{\Phi}_{\pm}(\vartheta)$ represent $MN$-dimensional vectors
		\begin{align}\label{eq3.7c}
			\boldsymbol{\Phi}_{\pm}(\vartheta) = \left[\boldsymbol{\phi}^{T}(\vartheta,\pm\mu_{1}),\: \boldsymbol{\phi}^{T}(\vartheta,\pm\mu_{2}),\: \dots,\: \boldsymbol{\phi}^{T}(\vartheta,\pm\mu_{N}) \right]^{T},
		\end{align}
		$\boldsymbol{D}$ 
		and $\boldsymbol{M}$ are diagonal matrices of order $MN$, such that,
		\begin{align}\label{eq3.7d}
			\boldsymbol{D} &= diag\left[\overbrace{\boldsymbol{\mathcal{T}},\: \boldsymbol{\mathcal{T}},\: \dots,\: \boldsymbol{\mathcal{T}}}^{N\,times}\right]\,,\\
			\boldsymbol{M} &= diag\left[ \mu_{1}\boldsymbol{I}_{M},\: \mu_{2}\boldsymbol{I}_{M},\: \dots,\: \mu_{N}\boldsymbol{I}_{M}\right],	\label{eq3.7e}
		\end{align}
		and $\boldsymbol{K}_{\pm}$ are square matrices of order $MN$ defined as
		\begin{align}\label{eq3.7f}
			\boldsymbol{K}_{\pm} = \left[\boldsymbol{L}^{\star}(\pm \mu_{1}),\: \boldsymbol{L}^{\star}(\pm \mu_{2}),\: \dots,\: \boldsymbol{L}^{\star}(\pm \mu_{N}) \right],
		\end{align}
		where $\boldsymbol{L}^{\star}(\mu_{n})$ is a $MN \times M$ matrix in the form 
		\begin{align}\label{eq3.7g}
			\boldsymbol{L}^{\star}(\mu_{n}) = \left[\overbrace{\boldsymbol{L}^{T}(\mu_{n})\omega_{n}, \: \boldsymbol{L}^{T}(\mu_{n})\omega_{n},\: \dots,\: \boldsymbol{L}^{T}(\mu_{n})\omega_{n}}^{N\,times} \right]^{T} .
		\end{align}
	\end{subequations}
	As described in \cref{s1}, we can reproduce the solution of the classical diffusion equation considering in the solution of the nonclassical transport equation the free-path distribution function as described in \cref{eq1.12a}. In this case, we have
	\begin{align}\label{eq3.8}
		\boldsymbol{K}_{+} = \boldsymbol{K}_{-} = \boldsymbol{K},
	\end{align}
	since $\mathcal{L}_{k}(\mu) = \mathcal{L}_{k}(-\mu) = \mathcal{L}_{k}$. 
	
	Now, we substitute \cref{eq3.8} into \cref{eq3.7a,eq3.7b} to obtain
	\begin{subequations}\label[pluraleq]{eq3.9}	
		\begin{align}\label{eq3.9a}
			\left(\boldsymbol{D} - \frac{1}{\vartheta}\boldsymbol{M} \right)\boldsymbol{\Phi}_{+}(\vartheta) = \frac{c}{2}\boldsymbol{K}\left[\boldsymbol{\Phi}_{+}(\vartheta) + \boldsymbol{\Phi}_{-}(\vartheta) \right] 
		\end{align}
		and
		\begin{align}\label{eq3.9b}
			\left(\boldsymbol{D} + \frac{1}{\vartheta}\boldsymbol{M} \right)\boldsymbol{\Phi}_{-}(\vartheta) = \frac{c}{2}\boldsymbol{K}\left[\boldsymbol{\Phi}_{+}(\vartheta) + \boldsymbol{\Phi}_{-}(\vartheta) \right].
		\end{align}	
	\end{subequations} 
	At this point, we follow two independent distinct procedures: (i) we sum up \cref{eq3.9a,eq3.9b}; and (ii) we subtract \cref{eq3.9b} from \cref{eq3.9a}. By doing these operations, we obtain
	\begin{subequations}\label[pluraleq]{eq3.10}	
		\begin{align}\label{eq3.10a}
			\left(\boldsymbol{D} - c\boldsymbol{K} \right)\boldsymbol{U}(\vartheta) = \frac{1}{\vartheta}\boldsymbol{M}\boldsymbol{V}(\vartheta)
		\end{align} 	
		and
		\begin{align}\label{eq3.10b}
			\boldsymbol{D}\boldsymbol{V}(\vartheta) = \frac{1}{\vartheta}\boldsymbol{M}\boldsymbol{U}(\vartheta),
		\end{align} 	
		where $\boldsymbol{U}(\vartheta)$ and $\boldsymbol{V}(\vartheta)$ are $MN$-dimensional vectors described as 
		\begin{align}\label{eq3.10c}
			\boldsymbol{U}(\vartheta) = \boldsymbol{\Phi}_{+}(\vartheta) + \boldsymbol{\Phi}_{-}(\vartheta)
		\end{align}	
		and
		\begin{align}\label{eq3.10d}
			\boldsymbol{V}(\vartheta) = \boldsymbol{\Phi}_{+}(\vartheta) - \boldsymbol{\Phi}_{-}(\vartheta).
		\end{align}	
	\end{subequations}
	Defining the $MN$-dimensional vectors $\boldsymbol{X}(\vartheta)$ and $\boldsymbol{Y}(\vartheta)$ as
	\begin{subequations}\label[pluraleq]{eq3.11}
		\begin{align}\label{eq3.11a}
			\boldsymbol{X}(\vartheta) = \boldsymbol{M}\boldsymbol{U}(\vartheta)
		\end{align}	
		and
		\begin{align}\label{eq3.11b}
			\boldsymbol{Y}(\vartheta) = \boldsymbol{M}\boldsymbol{V}(\vartheta),
		\end{align}		
	\end{subequations}
	we can obtain from \cref{eq3.10} the relations
	\begin{subequations}\label[pluraleq]{eq3.12}	
		\begin{align}\label{eq3.12a}
			\boldsymbol{A}\boldsymbol{X}(\vartheta) = \frac{1}{\vartheta}\boldsymbol{Y}(\vartheta)
		\end{align}
		and
		\begin{align}\label{eq3.12b}
			\boldsymbol{B}\boldsymbol{Y}(\vartheta) = \frac{1}{\vartheta}\boldsymbol{X}(\vartheta),
		\end{align}
		where $\boldsymbol{A}$ and $\boldsymbol{B}$ are square matrices of order $MN$, such that,
		\begin{align}\label{eq3.12c}
			\boldsymbol{A} = \left(\boldsymbol{D} - c\boldsymbol{K}\right)\boldsymbol{M}^{-1}
		\end{align}
		and
		\begin{align}\label{eq3.12d}
			\boldsymbol{B} =\boldsymbol{D}\boldsymbol{M}^{-1}.
		\end{align}
	\end{subequations}
	Finally, we use \cref{eq3.12b} to remove $\boldsymbol{Y}(\vartheta)$ from \cref{eq3.12a}, generating the equation
	\begin{align}\label{eq3.13}
		\boldsymbol{B}\boldsymbol{A}\boldsymbol{X}(\vartheta) = \frac{1}{\vartheta^{2}}\boldsymbol{X}(\vartheta),
	\end{align}
	which defines an eigenvalue problem of order $MN$. Solving \cref{eq3.13} we obtain $MN$ eigenvalues $(1/\vartheta^{2})$ and $MN$ eigenvectors $\boldsymbol{X}(\vartheta)$ of order $MN$. Thus, from an eigenvalue problem of order $MN$ we obtain $2MN$ constants $\pm \,\vartheta$. The eigenfunctions $\boldsymbol{\phi}$ can be calculated by the following relations
	\begin{subequations}\label[pluraleq]{eq3.14}
		\begin{align}\label{eq3.14a}
			\boldsymbol{\Phi}_{+}(\vartheta) = \frac{1}{2}\boldsymbol{M}^{-1}\left(\boldsymbol{I}_{MN} + \vartheta\boldsymbol{A} \right)\boldsymbol{X}(\vartheta) 
		\end{align}	
		and
		\begin{align}\label{eq3.14b}
			\boldsymbol{\Phi}_{-}(\vartheta) = \frac{1}{2}\boldsymbol{M}^{-1}\left(\boldsymbol{I}_{MN} - \vartheta\boldsymbol{A} \right)\boldsymbol{X}(\vartheta),
		\end{align}	
	\end{subequations}
	where $\boldsymbol{I}_{MN}$ represents the identity matrix of order $MN$. 
	
	Therefore, we can build the solution of \cref{eq3.2a} in the discrete ordinates formulation as a superposition of the solution proposed by \cref{eq3.4}. In other words, we have
	\begin{subequations}\label[pluraleq]{eq3.16}	
		\begin{align}\label{eq3.16a}
			\boldsymbol{\Psi}^{h}_{+}(x) = \sum_{j=1}^{MN}\left[\alpha_{j}\boldsymbol{\Phi}_{+}(\vartheta_{j})e^{-\frac{(x-x_{a})}{\vartheta_{j}}} + \beta_{j}\boldsymbol{\Phi}_{-}(\vartheta_{j})e^{-\frac{(x_{b}-x)}{\vartheta_{j}}}   \right] 
		\end{align}	
		and
		\begin{align}\label{eq3.16b}
			\boldsymbol{\Psi}^{h}_{-}(x) = \sum_{j=1}^{MN}\left[\alpha_{j}\boldsymbol{\Phi}_{-}(\vartheta_{j})e^{-\frac{(x-x_{a})}{\vartheta_{j}}} + \beta_{j}\boldsymbol{\Phi}_{+}(\vartheta_{j})e^{-\frac{(x_{b}-x)}{\vartheta_{j}}}   \right],
		\end{align}	
	\end{subequations}
	where $\alpha_{j}$ e $\beta_{j}$ are arbitrary constants and $\boldsymbol{\Psi}^{h}_{\pm}$ are $MN$-dimensional vectors defined as
	\begin{align}\label{eq3.17}
		\boldsymbol{\Psi}^{h}_{\pm}(x) = \left[\left( \boldsymbol{\psi}^{h}(x,\pm\mu_{1})\right)^{T} ,\: \left( \boldsymbol{\psi}^{h}(x,\pm\mu_{2})\right) ^{T},\: \dots,\: \left( \boldsymbol{\psi}^{h}(x,\pm\mu_{N})\right)^{T}\:  \right]^{T}.
	\end{align}
	In \cref{eq3.16} we have applied the exponential shift procedure \cite{Barichello:2011:Explicit} in order to avoid numerical overflows due to finite computational arithmetic. Thus, $x_{a}$ and $x_{b}$ represent the boundaries of the interval in which the homogeneous solution is defined.

	\subsection{Complex eigenvalues and eigenvectors}
	For complex eigenvalues it is convenient to write \cref{eq3.16} as presented in reference \cite{Neto:2018:Problema,NetoSubmitted}. Therefore, let us initially consider $\boldsymbol{\psi}^{h}_{1}$ as the homogeneous solution proposed in \cref{eq3.4} with $\vartheta_{j} = a_{j}+b_{j}\,i$, where $a_{j}$ and $b_{j}$ are positive numbers, such that,
	\begin{subequations}\label[pluraleq]{eq3.18}	
		\begin{align}\label{eq3.18a}
			\boldsymbol{\psi}^{h}_{1}(x,\mu) = \left[Re\left\lbrace \boldsymbol{\phi}(\vartheta_{j},\mu)\right\rbrace + i\,Im\left\lbrace \boldsymbol{\phi}(\vartheta_{j},\mu)\right\rbrace  \right]\left(cos\left(\frac{x\,b_{j}}{\vartheta_{j}\overline{\vartheta_{j}}} \right) +i\,sin\left(\frac{x\,b_{j}}{\vartheta_{j}\overline{\vartheta_{j}}} \right)   \right)e^{-\frac{x\,a_{j}}{\vartheta_{j}\overline{\vartheta_{j}}}},
		\end{align}
		where $Re$ and $Im$ are the real and imaginary parts of $\boldsymbol{\phi}$ and $\overline{\vartheta_{j}}$ is the complex conjugate of $\vartheta_{j}$. In \cref{eq3.18a} we used the Euler's formula to represent the complex exponential. As matrix $\boldsymbol{BA}$ in \cref{eq3.13} is real, complex eigenvalues always appear in conjugate pairs. This means that  
		\begin{align}\label{eq3.18b}
			\boldsymbol{\psi}^{h}_{2}(x,\mu) = \left[Re\left\lbrace \boldsymbol{\phi}(\overline{\vartheta_{j}},\mu)\right\rbrace + i\,Im\left\lbrace \boldsymbol{\phi}(\overline{\vartheta_{j}},\mu)\right\rbrace  \right]\left(cos\left(\frac{x\,b_{j}}{\vartheta_{j}\overline{\vartheta_{j}}} \right) -i\,sin\left(\frac{x\,b_{j}}{\vartheta_{j}\overline{\vartheta_{j}}} \right)  \right)e^{-\frac{x\,a_{j}}{\vartheta_{j}\overline{\vartheta_{j}}}}
		\end{align}
		is also a homogeneous solution.
		
		Analyzing the structure of the eigenfunctions presented in \cref{eq3.14} we note that
		\begin{align}\label{eq3.18c}
			\boldsymbol{\phi}(\overline{\vartheta_{j}},\mu) = \overline{\boldsymbol{\phi}(\vartheta_{j},\mu)}.
		\end{align} 
		Thus, we rewrite \cref{eq3.18b} using the property presented in \cref{eq3.18c}
		\begin{align}\label{eq3.18d}
			\boldsymbol{\psi}^{h}_{2}(x,\mu) = \left[Re\left\lbrace \boldsymbol{\phi}(\vartheta_{j},\mu)\right\rbrace - i\,Im\left\lbrace \boldsymbol{\phi}(\vartheta_{j},\mu)\right\rbrace  \right]\left(cos\left(\frac{x\,b_{j}}{\vartheta_{j}\overline{\vartheta_{j}}} \right) -i\,sin\left(\frac{x\,b_{j}}{\vartheta_{j}\overline{\vartheta_{j}}} \right)   \right)e^{-\frac{x\,a_{j}}{\vartheta_{j}\overline{\vartheta_{j}}}}.
		\end{align}
	\end{subequations}
	As $\boldsymbol{\psi}^{h}_{1}$ and $\boldsymbol{\psi}^{h}_{2}$ are homogeneous solutions, a superposition of these solutions is also a homogeneous solution. Therefore, we can build two real and linear independent solutions $\boldsymbol{\psi}^{h}_{1^{\star}}$ and  $\boldsymbol{\psi}^{h}_{2^{\star}}$ from the complex $\boldsymbol{\psi}^{h}_{1}$ and $\boldsymbol{\psi}^{h}_{2}$. Hence,
	\begin{subequations}\label[pluraleq]{eq3.19}	
		\begin{equation}\label{eq3.19a}
			\boldsymbol{\psi}^{h}_{1^{\star}}(x,\mu) = \left[Re\left\lbrace\boldsymbol{\phi}(\vartheta_{j},\mu) \right\rbrace cos\left(\frac{x\,b_{j}}{\vartheta_{j}\overline{\vartheta_{j}}} \right) - Im\left\lbrace\boldsymbol{\phi}(\vartheta_{j},\mu) \right\rbrace sin\left(\frac{x\,b_{j}}{\vartheta_{j}\overline{\vartheta_{j}}} \right)  \right]e^{-\frac{x\,a_{j}}{\vartheta_{j}\overline{\vartheta_{j}}}}
		\end{equation}
		and
		\begin{equation}\label{eq3.19b}
			\boldsymbol{\psi}^{h}_{2^{\star}}(x,\mu) = \left[Re\left\lbrace\boldsymbol{\phi}(\vartheta_{j},\mu) \right\rbrace sin\left(\frac{x\,b_{j}}{\vartheta_{j}\overline{\vartheta_{j}}} \right) + Im\left\lbrace\boldsymbol{\phi}(\vartheta_{j},\mu) \right\rbrace cos\left(\frac{x\,b_{j}}{\vartheta_{j}\overline{\vartheta_{j}}} \right)  \right]e^{-\frac{x\,a_{j}}{\vartheta_{j}\overline{\vartheta_{j}}}}.
		\end{equation}
	\end{subequations}
	
	Taking \cref{eq3.19} into consideration, we can rewrite \cref{eq3.16} for both real and complex eigenvalues and eigenvectors. That is \cite{Neto:2018:Problema,NetoSubmitted},
	\begin{subequations}\label[pluraleq]{eq3.20}	
		\begin{align}\label{eq3.20a}
			\boldsymbol{\Psi}^{h}_{+}(x) &= \sum_{j=1}^{J_{\mathcal{R}}}\left[\alpha_{j}\boldsymbol{\Phi}_{+}(\vartheta_{j})e^{-\frac{(x-x_{a})}{\vartheta_{j}}} + \beta_{j}\boldsymbol{\Phi}_{-}(\vartheta_{j})e^{-\frac{(x_{b}-x)}{\vartheta_{j}}} \right] +
			\\&
			\quad + \sum_{\begin{array}{c}
					j=J_{\mathcal{R}}+1 \\
					\Delta_{j}=2
			\end{array}}^{MN}\left\lbrace\left[\alpha_{j}\boldsymbol{H}^{+}_{1}((x-x_{a}),\vartheta_{j}) +\alpha_{j+1}\boldsymbol{H}^{+}_{2}((x-x_{a}),\vartheta_{j})  \right]e^{-\frac{(x-x_{a})}{\vartheta_{j}\overline{\vartheta_{j}}}} \right.+\nonumber
			\\&
			\qquad + \left. \left[\beta_{j}\boldsymbol{H}^{-}_{1}((x_{b}-x),\vartheta_{j}) +\beta_{j+1}\boldsymbol{H}^{-}_{2}((x_{b}-x),\vartheta_{j}) \right]e^{-\frac{(x_{b}-x)}{\vartheta_{j}\overline{\vartheta_{j}}}} \right\rbrace \nonumber
		\end{align}	
		and
		\begin{align}\label{eq3.20b}
			\boldsymbol{\Psi}^{h}_{-}(x) &= \sum_{j=1}^{J_{\mathcal{R}}}\left[\alpha_{j}\boldsymbol{\Phi}_{-}(\vartheta_{j})e^{-\frac{(x-x_{a})}{\vartheta_{j}}} + \beta_{j}\boldsymbol{\Phi}_{+}(\vartheta_{j})e^{-\frac{(x_{b}-x)}{\vartheta_{j}}} \right] + 
			\\&
			\quad + 
			\sum_{\begin{array}{c}
					j=J_{\mathcal{R}}+1 \\ [-0.1cm] \Delta_{j}=2
			\end{array}}^{MN}\left\lbrace\left[\alpha_{j}\boldsymbol{H}^{-}_{1}(x-x_{a},\vartheta_{j}) +\alpha_{j+1}\boldsymbol{H}^{-}_{2}(x-x_{a},\vartheta_{j})  \right]e^{-\frac{(x-x_{a})}{\vartheta_{j}\overline{\vartheta_{j}}}} \right.+ \nonumber
			\\&
			\qquad +
			\left. \left[\beta_{j}\boldsymbol{H}^{+}_{1}(x_{b}-x,\vartheta_{j}) +\beta_{j+1}\boldsymbol{H}^{+}_{2}(x_{b}-x,\vartheta_{j}) \right]e^{-\frac{(x_{b}-x)}{\vartheta_{j}\overline{\vartheta_{j}}}} \right\rbrace,\nonumber
		\end{align}
		where $J_{\mathcal{R}}$ represents the number of real and positive $\vartheta_{j}$ and $H^{\pm}_{1}$ and $H^{\pm}_{2}$ are $MN$-dimensional vectors defined as
		\begin{align}\label{eq3.20c}
			\boldsymbol{H}^{\pm}_{1}(x,\vartheta_{j}) = Re\left\lbrace\boldsymbol{\Phi}_{\pm}(\vartheta_{j}) \right\rbrace cos\left(\frac{x\,b_{j}}{\vartheta_{j}\overline{\vartheta_{j}}} \right) - Im\left\lbrace\boldsymbol{\Phi}_{\pm}(\vartheta_{j}) \right\rbrace sin\left(\frac{x\,b_{j}}{\vartheta_{j}\overline{\vartheta_{j}}} \right) 
		\end{align}
		and
		\begin{align}\label{eq3.20d}
			\boldsymbol{H}^{\pm}_{2}(x,\vartheta_{j}) = Re\left\lbrace\boldsymbol{\Phi}_{\pm}(\vartheta_{j}) \right\rbrace sin\left(\frac{x\,b_{j}}{\vartheta_{j}\overline{\vartheta_{j}}} \right) + Im\left\lbrace\boldsymbol{\Phi}_{\pm}(\vartheta_{j}) \right\rbrace cos\left(\frac{x\,b_{j}}{\vartheta_{j}\overline{\vartheta_{j}}} \right).
		\end{align}
	\end{subequations}
	\subsection{The general solution}\label{s3.2}
	After obtaining the homogeneous solution of 
	\cref{eq3.1a}, in the discrete ordinates formulation, we seek to find the particular solution of this equation in order to obtain the general solution.  Let us then consider a source $Q$ uniform with respect to the spatial variable inside the domain. Thus, we may assume that the particular solution will also be uniform with respect to the spatial variable. Therefore, we write \cref{eq3.1a} in convenient matrix form as
	\begin{align}\label{eq3.21}
		\boldsymbol{\mathcal{T}}\boldsymbol{\psi}^{p}(\mu) = \frac{c}{2}\int_{0}^{1}\left[\boldsymbol{L}(\mu')\boldsymbol{\psi}^{p}(\mu') + \boldsymbol{L}(-\mu')\boldsymbol{\psi}^{p}(-\mu') \right]d\mu' + \boldsymbol{Q},
	\end{align}
	where $\boldsymbol{Q}$ and $\boldsymbol{\psi}^{p}(\mu)$ are $M$-dimensional vectors whose entries are $Q$ and $\psi^{p}_{m}(\mu)$ respectively, with $\psi^{p}_{m}(\mu)$ representing the particular solution of \cref{eq3.1a}.
	As with the homogeneous solution, we consider the same quadrature scheme defined in the semi-interval $[0, 1]$, to rewrite \cref{eq3.21} as a linear system composed of $2MN$ equations. That is,
	\begin{subequations}\label[pluraleq]{eq3.22}
		\begin{align}\label{eq3.22a}
			\boldsymbol{\mathcal{T}}\boldsymbol{\psi}^{p}(\mu_{n}) = \frac{c}{2}\sum_{i=1}^{N}\left[\boldsymbol{L}(\mu_{i})\boldsymbol{\psi}^{p}(\mu_{i}) + \boldsymbol{L}(-\mu_{i})\boldsymbol{\psi}^{p}(-\mu_{i}) \right]\omega_{i} + \boldsymbol{Q}
		\end{align}
		and
		\begin{align}\label{eq3.22b}
			\boldsymbol{\mathcal{T}}\boldsymbol{\psi}^{p}(-\mu_{n}) = \frac{c}{2}\sum_{i=1}^{N}\left[\boldsymbol{L}(\mu_{i})\boldsymbol{\psi}^{p}(\mu_{i}) + \boldsymbol{L}(-\mu_{i})\boldsymbol{\psi}^{p}(-\mu_{i}) \right]\omega_{i} + \boldsymbol{Q},
		\end{align}
	\end{subequations}
	where $n=1,2,\dots,N$. From \cref{eq3.22} we conclude
	\begin{align}\label{eq3.23}
		\boldsymbol{\psi}^{p}(\mu_{n}) = \boldsymbol{\psi}^{p}(-\mu_{n}).
	\end{align}
	Furthermore, varying $n$ from 1 to $N$ in \cref{eq3.22a} and making use of the relations presented in \cref{eq3.8,eq3.23}, we obtain
	\begin{subequations}\label[pluraleq]{eq3.24}
		\begin{align}\label{eq3.24a}
			\boldsymbol{D}\boldsymbol{\Psi}^{p}_{\pm} = c\boldsymbol{K}\boldsymbol{\Psi}^{p}_{\pm} + \boldsymbol{S},
		\end{align}
		where $\boldsymbol{\Psi}^{p}_{\pm}$ is a vector of order $MN$ defined as
		\begin{align}\label{eq3.24b}
			\boldsymbol{\Psi}^{p}_{\pm} = \left[\left( \boldsymbol{\psi}^{p}(\pm\mu_{1})\right)^{T} ,\:\left( \boldsymbol{\psi}^{p}(\pm\mu_{2})\right)^{T} ,\: \dots,\: \left( \boldsymbol{\psi}^{p}(\pm\mu_{N})\right)^{T}\,  \right]^{T} 
		\end{align}
		and $\boldsymbol{S}$ is a vector of order $MN$ composed by vector $\boldsymbol{Q}$ repeated $N$ times.
		The particular solutions can be obtained from \cref{eq3.24a} as
	\end{subequations}
	\begin{align}\label{eq3.25}
		\boldsymbol{\Psi}^{p}_{\pm} = \left(\boldsymbol{D}-c\boldsymbol{K}\right)^{-1}\boldsymbol{S},
	\end{align}
	provided matrix $\left(\boldsymbol{D}-c\boldsymbol{K}\right)$ is non singular.
	
	Having found the homogeneous and particular solutions, we can write the general solution in the following closed form
	\begin{subequations}\label[pluraleq]{eq3.26}	
		\begin{align}\label{eq3.26a}
			\boldsymbol{\Psi}_{+}(x) &= \sum_{j=1}^{J_{\mathcal{R}}}\left[\alpha_{j}\boldsymbol{\Phi}_{+}(\vartheta_{j})e^{-\frac{(x-x_{a})}{\vartheta_{j}}} + \beta_{j}\boldsymbol{\Phi}_{-}(\vartheta_{j})e^{-\frac{(x_{b}-x)}{\vartheta_{j}}} \right] + 
			\\&
			\quad +
			\sum_{\begin{array}{c}
					j=J_{\mathcal{R}}+1 \\ [-0.1cm] \Delta_{j}=2
			\end{array}}^{MN}\left\lbrace\left[\alpha_{j}\boldsymbol{H}^{+}_{1}(x-x_{a},\vartheta_{j}) +\alpha_{j+1}\boldsymbol{H}^{+}_{2}(x-x_{a},\vartheta_{j})  \right]e^{-\frac{(x-x_{a})}{\vartheta_{j}\overline{\vartheta_{j}}}} \right.+ \nonumber
			\\&
			\qquad +	\left. \left[\beta_{j}\boldsymbol{H}^{-}_{1}(x_{b}-x,\vartheta_{j}) +\beta_{j+1,r}\boldsymbol{H}^{-}_{2}(x_{b}-x,\vartheta_{j}) \right]e^{-\frac{(x_{b}-x)}{\vartheta_{j}\overline{\vartheta_{j}}}} \right\rbrace + \boldsymbol{\Psi}^{p}_{+} \nonumber
		\end{align}	
		and
		\begin{align}\label{eq3.26b}
			\boldsymbol{\Psi}_{-}(x) &= \sum_{j=1}^{J_{\mathcal{R}}}\left[\alpha_{j}\boldsymbol{\Phi}_{-}(\vartheta_{j})e^{-\frac{(x-x_{a})}{\vartheta_{j}}} + \beta_{j}\boldsymbol{\Phi}_{+}(\vartheta_{j})e^{-\frac{(x_{b}-x)}{\vartheta_{j}}} \right] + 
			\\&
			\quad +
			\sum_{\begin{array}{c}
					j=J_{\mathcal{R}}+1 \\ [-0.1cm] \Delta_{j}=2
			\end{array}}^{MN}\left\lbrace\left[\alpha_{j}\boldsymbol{H}^{-}_{1}(x-x_{a},\vartheta_{j}) +\alpha_{j+1}\boldsymbol{H}^{-}_{2}(x-x_{a},\vartheta_{j})  \right]e^{-\frac{(x-x_{a})}{\vartheta_{j}\overline{\vartheta_{j}}}} \right.+\nonumber
			\\&
			\qquad +
			\left. \left[\beta_{j}\boldsymbol{H}^{+}_{1}(x_{b}-x,\vartheta_{j}) +\beta_{j+1}\boldsymbol{H}^{+}_{2}(x_{b}-x,\vartheta_{j}) \right]e^{-\frac{(x_{b}-x)}{\vartheta_{j}\overline{\vartheta_{j}}}} \right\rbrace + \boldsymbol{\Psi}^{p}_{-}, \nonumber
		\end{align}
	\end{subequations}
	where $\boldsymbol{\Psi}^{p}_ {\pm}$ are given by \cref{eq3.25}. In order to fully establish the general solution, we must determine the constants $\alpha_{j}$ and $\beta_{j}$. The arbitrary constants are obtained through the solution of a linear system of order $2MN$ generated by the boundary conditions (\cref{eq3.1b,eq3.1c}) and \cref{eq3.26}. 
	
	In the next section we perform numerical experiments considering that the source $Q$ can vary its intensity along different regions of the domain. However, $Q$ is still uniform with respect to the spatial variable within these regions. In this case, we must apply the ADO method in each region to obtain the general solution for the problem. Therefore, $2MNR$ arbitrary constants are generated, where $R$ represents the number of regions in which the source $Q$ varies its intensity. To determine the arbitrary constants and completely establish the solution of this problem, we generate and solve a linear system of order $2MNR$ making use of the boundary conditions (\cref{eq3.1b,eq3.1c}) and the continuity conditions
	\begin{align}\label{eq3.27}
		\psi_{m,r}(x_{r}) = \psi_{m,r+1}(x_{r}), \: r = 1,2,\dots,R-1,
	\end{align}
	where $\psi_{m,r}$ and $\psi_{m,r+1}$ represent the local general solutions obtained in two adjacent regions with $x_{r}$ being the intersection point of these regions.
	
	\section{Numerical results}\label{s4}
	\setcounter{equation}{0}
	In this section we present numerical results for two test problems, with the aim of describing in detail some challenges that may arise from the use of the spectral approximation, and analyzing the performance of the ADO method in solving \cref{eq3.1a}. 
	To achieve this goal, we reproduce the solution of the one-dimensional classical diffusion equation
	\begin{align}\label{eq4.1}
		-\f{1}{3\sigma_t}\f{d^2}{dx^2}\Phi(x) + (1-c)\sigma_t\Phi(x) = Q(x)
	\end{align}
	by solving the equivalent nonclassical transport problem
	\begin{align}\label{eq4.2}
		\Phi(x) = \frac{1}{\sigma_{t}}\sum_{k=0}^{M}\mathcal{L}_{k}\sum_{n=1}^{N}\left[ \psi_{k}(x,\mu_{n})+\psi_{k}(x,-\mu_{n})\right] \omega_{n},
	\end{align}
	where the functions $\mathcal{L}_{k}$ are given by \cref{eq2.14}. We consider vacuum boundary conditions as given by \cref{eq3.1b,eq3.1c}, and (when needed) continuity conditions as described in \cref{eq3.27}. Moreover, Gauss-Legendre angular quadratures, mapped to the half-range [0,1], are considered. The nonclassical solution given in \cref{eq4.2} is compared with the solution of \cref{eq4.1} with Mark (Vacuum) boundary conditions. The solution of \cref{eq4.1} was implemented following the procedure described in reference \cite{duderstadtsol}. Thus, we calculate the homogeneous and particular solutions that compose the analytic general solution of \cref{eq4.1}, and then use the boundary conditions and (when needed) continuity conditions to determine the arbitrary constants.
	
	As discussed in \cref{s1}, the appropriate way to define boundary conditions to the nonclassical transport equation is not universally agreed upon. Therefore, the exact correlation between the boundary conditions considered for the solutions of \cref{eq4.1,eq4.2} is not completely clear, and need further investigation. We chose to use Mark boundary conditions due to its greater performance, for the discrete ordinates models considered in this work, compared to other standard diffusion boundary conditions.

	\subsection{Test Problem 1}
	Let us consider a slab of length $X = 20\,cm$, with $\sigma_{t} = 0.578\,cm^{-1}$.
	In this system, we introduce an isotropic source $Q$, such that
	\begin{align}\label{eq4.3}
		Q(x) = \left\lbrace \begin{array}{l}
			1,\: x_{1} \leq x \leq x_{2}, \\
			0,\: \text{otherwise}
		\end{array} \right..
	\end{align} 
	The choice of $x_{1}$ and $x_{2}$ will define the interval upon which the source emits particles.
	
	\Cref{p11,p12,p13} present solutions of \cref{eq4.1,eq4.2} for scattering ratios $c = 0.3$, $c = 0.9$, and $c = 0.99$, respectively.
	The relative errors of the nonclassical transport solution with respect to the analytical solution of the diffusion problem are also given.
	In all cases, the source $Q$, as defined by \cref{eq4.3}, is located at the center of the slab, with boundaries $x_{1} = 9.5\,cm$ and $x_{2} = 10.5\,cm$.
	\begin{table}[H]
		\caption{Neutron scalar flux for Test Problem 1, with $c = 0.3$, $x_{1}=9.5\,cm$, and $x_{2} = 10.5\,cm$.}
		\begin{center}
			{\renewcommand{\arraystretch}{1.15}
				\adjustbox{angle=90}{
					\begin{threeparttable}
						\begin{tabular}{cccccc|cccc}
							\toprule 
							$x^{\star}$\tnote{a} & Solution of \cref{eq4.1} & \multicolumn{4}{c}{Nonclassical Solution (\cref{eq4.2})} & \multicolumn{4}{c}{Relative Error}\tabularnewline
							$(cm)$ & $(neutrons/cm^{2}s)$ & \multicolumn{4}{c}{$(neutrons/cm^{2}s)$} & \multicolumn{4}{c}{}\tabularnewline
							\cmidrule{3-10} \cmidrule{4-10} \cmidrule{5-10} \cmidrule{6-10} \cmidrule{7-10} \cmidrule{8-10} \cmidrule{9-10} \cmidrule{10-10} 
							&  & $N = 20$ & $N = 40$ & $N = 60$ & $N = 80$ & $N = 20$ & $N = 40$ & $N = 60$ & $N = 80$\tabularnewline
							\cmidrule{3-10} \cmidrule{4-10} \cmidrule{5-10} \cmidrule{6-10} \cmidrule{7-10} \cmidrule{8-10} \cmidrule{9-10} \cmidrule{10-10} 
							&  & \multicolumn{8}{c}{$M = 1$}\tabularnewline
							\midrule
							0.0 & 1.691372E+00\tnote{b} & 1.697259E+00 & 1.692122E+00 & 1.692477E+00 & 1.692465E+00 & 3.4E-03 & 4.4E-04 & 6.5E-04 & 6.4E-04\tabularnewline
							2.0 & 3.991187E-01 & 3.989473E-01 & 3.989748E-01 & 3.989743E-01 & 3.989743E-01 & 4.2E-04 & 3.6E-04 & 3.6E-04 & 3.6E-04\tabularnewline
							4.0 & 7.474274E-02 & 7.480994E-02 & 7.480989E-02 & 7.480989E-02 & 7.480989E-02 & 8.9E-04 & 8.9E-04 & 8.9E-04 & 8.9E-04\tabularnewline
							6.0 & 1.399560E-02 & 1.402761E-02 & 1.402761E-02 & 1.402761E-02 & 1.402761E-02 & 2.2E-03 & 2.2E-03 & 2.2E-03 & 2.2E-03\tabularnewline
							8.0 & 2.613071E-03 & 2.622733E-03 & 2.622736E-03 & 2.622737E-03 & 2.622737E-03 & 3.6E-03 & 3.6E-03 & 3.6E-03 & 3.6E-03\tabularnewline
							10.0 & 4.472256E-04 & 4.496045E-04 & 4.495229E-04 & 4.495074E-04 & 4.495019E-04 & 5.3E-03 & 5.1E-03 & 5.1E-03 & 5.0E-03\tabularnewline
							\midrule 
							&  & \multicolumn{8}{c}{$M = 2$}\tabularnewline
							\midrule
							0.0 & 1.691372E+00 & 1.696157E+00 & 1.691037E+00 & 1.691392E+00 & 1.691379E+00 & 2.8E-03 & 1.9E-04 & 1.1E-05 & 3.7E-06\tabularnewline
							2.0 & 3.991187E-01 & 3.990916E-01 & 3.991193E-01 & 3.991188E-01 & 3.991188E-01 & 6.7E-05 & 1.5E-06 & 3.2E-07 & 3.5E-07\tabularnewline
							4.0 & 7.474274E-02 & 7.474287E-02 & 7.474282E-02 & 7.474282E-02 & 7.474282E-02 & 1.7E-06 & 1.0E-06 & 1.0E-06 & 1.0E-06\tabularnewline
							6.0 & 1.399560E-02 & 1.399559E-02 & 1.399559E-02 & 1.399559E-02 & 1.399559E-02 & 5.2E-07 & 6.9E-07 & 6.8E-07 & 6.8E-07\tabularnewline
							8.0 & 2.613071E-03 & 2.613056E-03 & 2.613058E-03 & 2.613059E-03 & 2.613059E-03 & 5.7E-06 & 4.7E-06 & 4.4E-06 & 4.4E-06\tabularnewline
							10.0 & 4.472256E-04 & 4.473307E-04 & 4.472494E-04 & 4.472339E-04 & 4.472284E-04 & 2.3E-04 & 5.3E-05 & 1.8E-05 & 6.4E-06\tabularnewline
							\midrule 
							&  & \multicolumn{8}{c}{$M = 3$}\tabularnewline
							\midrule
							0.0 & 1.691372E+00 & 1.696156E+00 & 1.691036E+00 & 1.691391E+00 & 1.691378E+00 & 2.8E-03 & 1.9E-04 & 1.1E-05 & 3.3E-06\tabularnewline
							2.0 & 3.991187E-01 & 3.990914E-01 & 3.991191E-01 & 3.991187E-01 & 3.991187E-01 & 6.8E-05 & 1.1E-06 & 4.4E-08 & 1.5E-08\tabularnewline
							4.0 & 7.474274E-02 & 7.474278E-02 & 7.474273E-02 & 7.474273E-02 & 7.474273E-02 & 6.0E-07 & 6.9E-08 & 6.7E-08 & 6.7E-08\tabularnewline
							6.0 & 1.399560E-02 & 1.399560E-02 & 1.399560E-02 & 1.399560E-02 & 1.399560E-02 & 4.1E-08 & 1.3E-07 & 1.2E-07 & 1.2E-07\tabularnewline
							8.0 & 2.613071E-03 & 2.613066E-03 & 2.613069E-03 & 2.613070E-03 & 2.613070E-03 & 1.6E-06 & 5.6E-07 & 3.5E-07 & 2.7E-07\tabularnewline
							10.0 & 4.472256E-04 & 4.473347E-04 & 4.472534E-04 & 4.472379E-04 & 4.472325E-04 & 2.4E-04 & 6.2E-05 & 2.7E-05 & 1.5E-05\tabularnewline
							\bottomrule
						\end{tabular}
						\begin{tablenotes}[flushleft]
							\footnotesize
							\item[a] $x = \pm x^{\star}+10.0$. For example, if $x^{\star} = 2.0$, the results presented are valid for $x = 8.0$ and $x = 12.0$. We use $x^{\star}$ due to the problem's symmetry at $x = 10.0$.
							\item[b] Read as 1.691372$\times10^{+00}$.
						\end{tablenotes}
					\end{threeparttable}
				}
			}
		\end{center}
		\label{p11}
	\end{table}
	
	\begin{table}[H]
		\caption{Neutron scalar flux for Test Problem 1, with $c = 0.9$, $x_{1}=9.5\,cm$, and $x_{2} = 10.5\,cm$.}
		\begin{center}
			{\renewcommand{\arraystretch}{1.15}
				\adjustbox{angle=90}{
					\begin{threeparttable}
						\begin{tabular}{cccccc|cccc}
							\toprule 
							$x^{\star}$\tnote{a} & Solution of \cref{eq4.1} & \multicolumn{4}{c}{Nonclassical Solution (\cref{eq4.2})} & \multicolumn{4}{c}{Relative Error}\tabularnewline
							$(cm)$ & $(neutrons/cm^{2}s)$ & \multicolumn{4}{c}{$(neutrons/cm^{2}s)$} & \multicolumn{4}{c}{}\tabularnewline
							\cmidrule{3-10} \cmidrule{4-10} \cmidrule{5-10} \cmidrule{6-10} \cmidrule{7-10} \cmidrule{8-10} \cmidrule{9-10} \cmidrule{10-10} 
							&  & $N = 20$ & $N = 40$ & $N = 60$ & $N = 80$ & $N = 20$ & $N = 40$ & $N = 60$ & $N = 80$\tabularnewline
							\cmidrule{3-10} \cmidrule{4-10} \cmidrule{5-10} \cmidrule{6-10} \cmidrule{7-10} \cmidrule{8-10} \cmidrule{9-10} \cmidrule{10-10} 
							&  & \multicolumn{8}{c}{$M = 1$}\tabularnewline
							\midrule
							0.0 & 5.055564E+00\tnote{b} & 5.061230E+00 & 5.055958E+00 & 5.056317E+00 & 5.056304E+00 & 1.1E-03 & 7.7E-05 & 1.4E-04 & 1.4E-04\tabularnewline
							2.0 & 2.907797E+00 & 2.906954E+00 & 2.906984E+00 & 2.906985E+00 & 2.906985E+00 & 2.8E-04 & 2.7E-04 & 2.7E-04 & 2.7E-04\tabularnewline
							4.0 & 1.530832E+00 & 1.530778E+00 & 1.530786E+00 & 1.530788E+00 & 1.530788E+00 & 3.5E-05 & 2.9E-05 & 2.8E-05 & 2.8E-05\tabularnewline
							6.0 & 7.883585E-01 & 7.885191E-01 & 7.885335E-01 & 7.885362E-01 & 7.885372E-01 & 2.0E-04 & 2.2E-04 & 2.2E-04 & 2.2E-04\tabularnewline
							8.0 & 3.726391E-01 & 3.727884E-01 & 3.728148E-01 & 3.728198E-01 & 3.728216E-01 & 4.0E-04 & 4.7E-04 & 4.8E-04 & 4.8E-04\tabularnewline
							10.0 & 1.113692E-01 & 1.114802E-01 & 1.114572E-01 & 1.114529E-01 & 1.114513E-01 & 9.9E-04 & 7.9E-04 & 7.5E-04 & 7.3E-04\tabularnewline
							\midrule 
							&  & \multicolumn{8}{c}{$M = 2$}\tabularnewline
							\midrule
							0.0 & 5.055564E+00 & 5.060477E+00 & 5.055224E+00 & 5.055582E+00 & 5.055569E+00 & 9.7E-04 & 6.7E-05 & 3.5E-06 & 1.0E-06\tabularnewline
							2.0 & 2.907797E+00 & 2.907765E+00 & 2.907796E+00 & 2.907796E+00 & 2.907796E+00 & 1.0E-05 & 4.5E-07 & 2.8E-07 & 1.6E-07\tabularnewline
							4.0 & 1.530832E+00 & 1.530822E+00 & 1.530830E+00 & 1.530831E+00 & 1.530832E+00 & 6.7E-06 & 1.5E-06 & 5.7E-07 & 2.2E-07\tabularnewline
							6.0 & 7.883585E-01 & 7.883394E-01 & 7.883537E-01 & 7.883565E-01 & 7.883574E-01 & 2.4E-05 & 6.0E-06 & 2.6E-06 & 1.3E-06\tabularnewline
							8.0 & 3.726391E-01 & 3.726037E-01 & 3.726300E-01 & 3.726351E-01 & 3.726368E-01 & 9.5E-05 & 2.4E-05 & 1.0E-05 & 6.1E-06\tabularnewline
							10.0 & 1.113692E-01 & 1.114001E-01 & 1.113771E-01 & 1.113727E-01 & 1.113712E-01 & 2.7E-04 & 7.0E-05 & 3.1E-05 & 1.7E-05\tabularnewline
							\midrule 
							&  & \multicolumn{8}{c}{$M = 3$}\tabularnewline
							\midrule
							0.0 & 5.055564E+00 & 5.060476E+00 & 5.055223E+00 & 5.055582E+00 & 5.055569E+00 & 9.7E-04 & 6.7E-05 & 3.5E-06 & 1.0E-06\tabularnewline
							2.0 & 2.907797E+00 & 2.907765E+00 & 2.907796E+00 & 2.907796E+00 & 2.907796E+00 & 1.0E-05 & 4.3E-07 & 2.7E-07 & 1.5E-07\tabularnewline
							4.0 & 1.530832E+00 & 1.530821E+00 & 1.530829E+00 & 1.530831E+00 & 1.530831E+00 & 7.0E-06 & 1.8E-06 & 8.1E-07 & 4.6E-07\tabularnewline
							6.0 & 7.883585E-01 & 7.883392E-01 & 7.883536E-01 & 7.883563E-01 & 7.883572E-01 & 2.4E-05 & 6.2E-06 & 2.8E-06 & 1.6E-06\tabularnewline
							8.0 & 3.726391E-01 & 3.726036E-01 & 3.726300E-01 & 3.726350E-01 & 3.726368E-01 & 9.5E-05 & 2.4E-05 & 1.1E-05 & 6.2E-06\tabularnewline
							10.0 & 1.113692E-01 & 1.114001E-01 & 1.113771E-01 & 1.113727E-01 & 1.113712E-01 & 2.7E-04 & 7.0E-05 & 3.1E-05 & 1.7E-05\tabularnewline
							\bottomrule
						\end{tabular}
						\begin{tablenotes}[flushleft]
							\footnotesize
							\item[a] $x = \pm x^{\star} +10.0$. For example, if $x^{\star} = 2.0$, the results presented are valid for $x = 8.0$ and $x = 12.0$.
							\item[b] Read as 5.055564$\times10^{+00}$.
						\end{tablenotes}
					\end{threeparttable}
				}
			}
		\end{center}
		\label{p12}
	\end{table}
	
	\begin{table}[H]
		\caption{Neutron scalar flux for Test Problem 1, with $c = 0.99$, $x_{1}=9.5\,cm$, and $x_{2} = 10.5\,cm$.}
		\begin{center}
			{\renewcommand{\arraystretch}{1.15}
				\adjustbox{angle=90}{
					\begin{threeparttable}
						\begin{tabular}{cccccc|cccc}
							\toprule 
							$x^{\star}$\tnote{a} & Solution of \cref{eq4.1} & \multicolumn{4}{c}{Nonclassical Solution (\cref{eq4.2})} & \multicolumn{4}{c}{Relative Error}\tabularnewline
							$(cm)$ & $(neutrons/cm^{2}s)$ & \multicolumn{4}{c}{$(neutrons/cm^{2}s)$} & \multicolumn{4}{c}{}\tabularnewline
							\cmidrule{3-10} \cmidrule{4-10} \cmidrule{5-10} \cmidrule{6-10} \cmidrule{7-10} \cmidrule{8-10} \cmidrule{9-10} \cmidrule{10-10} 
							&  & $N = 20$ & $N = 40$ & $N = 60$ & $N = 80$ & $N = 20$ & $N = 40$ & $N = 60$ & $N = 80$\tabularnewline
							\cmidrule{3-10} \cmidrule{4-10} \cmidrule{5-10} \cmidrule{6-10} \cmidrule{7-10} \cmidrule{8-10} \cmidrule{9-10} \cmidrule{10-10} 
							&  & \multicolumn{8}{c}{$M = 1$}\tabularnewline
							\midrule
							0.0 & 1.344633E+01\tnote{b} & 1.344684E+01 & 1.344199E+01 & 1.344243E+01 & 1.344245E+01 & 3.7E-05 & 3.2E-04 & 2.8E-04 & 2.8E-04\tabularnewline
							2.0 & 1.066639E+01 & 1.066063E+01 & 1.066111E+01 & 1.066119E+01 & 1.066123E+01 & 5.4E-04 & 4.9E-04 & 4.8E-04 & 4.8E-04\tabularnewline
							4.0 & 7.881924E+00 & 7.877921E+00 & 7.878405E+00 & 7.878496E+00 & 7.878529E+00 & 5.0E-04 & 4.4E-04 & 4.3E-04 & 4.3E-04\tabularnewline
							6.0 & 5.414493E+00 & 5.411868E+00 & 5.412398E+00 & 5.412499E+00 & 5.412534E+00 & 4.8E-04 & 3.8E-04 & 3.6E-04 & 3.6E-04\tabularnewline
							8.0 & 3.164856E+00 & 3.163337E+00 & 3.163932E+00 & 3.164046E+00 & 3.164086E+00 & 4.8E-04 & 2.9E-04 & 2.5E-04 & 2.4E-04\tabularnewline
							10.0 & 1.042522E+00 & 1.042478E+00 & 1.042418E+00 & 1.042406E+00 & 1.042402E+00 & 4.2E-05 & 1.0E-04 & 1.1E-04 & 1.1E-04\tabularnewline
							\midrule 
							&  & \multicolumn{8}{c}{$M = 2$}\tabularnewline
							\midrule
							0.0 & 1.344633E+01 & 1.345066E+01 & 1.344583E+01 & 1.344627E+01 & 1.344629E+01 & 3.2E-04 & 3.6E-05 & 3.8E-06 & 2.5E-06\tabularnewline
							2.0 & 1.066639E+01 & 1.066576E+01 & 1.066624E+01 & 1.066632E+01 & 1.066635E+01 & 5.9E-05 & 1.4E-05 & 6.7E-06 & 3.8E-06\tabularnewline
							4.0 & 7.881924E+00 & 7.881274E+00 & 7.881758E+00 & 7.881849E+00 & 7.881882E+00 & 8.2E-05 & 2.1E-05 & 9.5E-06 & 5.3E-06\tabularnewline
							6.0 & 5.414493E+00 & 5.413780E+00 & 5.414310E+00 & 5.414411E+00 & 5.414446E+00 & 1.3E-04 & 3.3E-05 & 1.5E-05 & 8.6E-06\tabularnewline
							8.0 & 3.164856E+00 & 3.164054E+00 & 3.164650E+00 & 3.164763E+00 & 3.164803E+00 & 2.5E-04 & 6.5E-05 & 2.9E-05 & 1.6E-05\tabularnewline
							10.0 & 1.042522E+00 & 1.042603E+00 & 1.042543E+00 & 1.042531E+00 & 1.042527E+00 & 7.7E-05 & 1.9E-05 & 8.5E-06 & 4.6E-06\tabularnewline
							\midrule 
							&  & \multicolumn{8}{c}{$M = 3$}\tabularnewline
							\midrule
							0.0 & 1.344633E+01 & 1.345066E+01 & 1.344583E+01 & 1.344628E+01 & 1.344629E+01 & 3.2E-04 & 3.6E-05 & 3.7E-06 & 2.4E-06\tabularnewline
							2.0 & 1.066639E+01 & 1.066576E+01 & 1.066624E+01 & 1.066632E+01 & 1.066635E+01 & 6.0E-05 & 1.4E-05 & 6.6E-06 & 3.7E-06\tabularnewline
							4.0 & 7.881924E+00 & 7.881274E+00 & 7.881758E+00 & 7.881849E+00 & 7.881882E+00 & 8.2E-05 & 2.1E-05 & 9.4E-06 & 5.3E-06\tabularnewline
							6.0 & 5.414493E+00 & 5.413780E+00 & 5.414311E+00 & 5.414411E+00 & 5.414447E+00 & 1.3E-04 & 3.3E-05 & 1.5E-05 & 8.5E-06\tabularnewline
							8.0 & 3.164856E+00 & 3.164054E+00 & 3.164650E+00 & 3.164763E+00 & 3.164803E+00 & 2.5E-04 & 6.5E-05 & 2.9E-05 & 1.6E-05\tabularnewline
							10.0 & 1.042522E+00 & 1.042603E+00 & 1.042542E+00 & 1.042531E+00 & 1.042527E+00 & 7.7E-05 & 1.9E-05 & 8.4E-06 & 4.5E-06\tabularnewline
							\bottomrule
						\end{tabular}
						\begin{tablenotes}[flushleft]
							\footnotesize
							\item[a] $x = \pm x^{\star} +10.0$. For example, if $x^{\star} = 2.0$, the results presented are valid for $x = 8.0$ and $x = 12.0$.
							\item[b] Read as 1.344633$\times10^{+01}$.
						\end{tablenotes}
					\end{threeparttable}
				}
			}
		\end{center}
		\label{p13}
	\end{table}
	
	As can be seen in \cref{p11,p12,p13}, the ADO method produces accurate results for Test Problem 1, with precision increasing as $N$ and $M$ increase.
	As expected, the best results are obtained when using $N=80$ and $M=3$, since this choice of parameters offers a better representation of the scalar flux generated by using \cref{eq4.2}.
	For all the cases, there is agreement up to at least 5 decimal places when varying $N$ from 60 to 80, and up to at least 6 decimal places when varying $M$ from 2 to 3.
	Thus, to obtain a solution accurate to 5 decimal places, the choice of parameters $N=60$ and $M=2$ would suffice for this model problem.
	This is confirmed when analyzing the maximum relative errors displayed in \cref{p11,p12,p13} for these choices of $N$ and $M$, which is $3.2\times 10^{-05}$.
	We remark that, in order to obtain results with higher precision, the values of $N$ and $M$ in \cref{eq4.2} would need to be larger.
	
	It is also noticeable that the ADO method does not seem to be too sensitive to changes in the scattering ratio $c$.
	For instance, when analyzing the absolute relative deviations obtained with $N=60$ and $M=2$, the loss in accuracy observed when $c$ increases is very small.
	This indicates, in this case, that the matrices built by the ADO method are well-conditioned.
	
	Next, we will allow the isotropic source $Q$ to emit neutrons in the whole domain.
	Using the same choices of cross section $\sigma_t$ and scattering ratios $c$ used in the previous examples, \cref{p14,p15,p16} display the solutions of \cref{eq4.1,eq4.2} for Test Problem 1 as the scattering ratio $c$ increases, with $x_{1} = 0\,cm$ and $x_{2} = 20\,cm$ in \cref{eq4.3}.
	We also show the relative errors of the nonclassical transport solution obtained with the ADO method when compared to the analytical solution of \cref{eq4.1}. 
	
	The ADO method also generates accurate results for problems with a uniform source in the whole domain.
	Once again, as expected, the best results occur when $N=80$ and $M=3$, with agreement between 5 and 7 decimal places with respect to the analytical solution of \cref{eq4.1}.
	As in the previous results for Test Problem 1, solutions of \cref{eq4.2} obtained when varying $N$ from 60 to 80 show agreement between 5 and 6 decimal places.
	Similarly, there is agreement between 6 and 7 decimal places in the solutions of \cref{eq4.2} attained when varying $M$ from 2 to 3. 
	This follows the trend of the observation made previously for the results presented in \cref{p11,p12,p13}, that choosing $N=60$ and $M=2$ should suffice when searching for a solution of \cref{eq4.2} that is accurate to 5 decimal places.

	\begin{table}[]
		\caption{Neutron scalar flux for Test Problem 1, with $c = 0.3$, $x_{1}=0.0\,cm$, and $x_{2} = 20.0\,cm$.}
		\begin{center}
			{\renewcommand{\arraystretch}{1.15}
				\adjustbox{angle=90}{
					\begin{threeparttable}
						\begin{tabular}{cccccc|cccc}
							\toprule 
							$x^{\star}$\tnote{a} & Solution of \cref{eq4.1} & \multicolumn{4}{c}{Nonclassical Solution (\cref{eq4.2})} & \multicolumn{4}{c}{Relative Error}\tabularnewline
							$(cm)$ & $(neutrons/cm^{2}s)$ & \multicolumn{4}{c}{$(neutrons/cm^{2}s)$} & \multicolumn{4}{c}{}\tabularnewline
							\cmidrule{3-10} \cmidrule{4-10} \cmidrule{5-10} \cmidrule{6-10} \cmidrule{7-10} \cmidrule{8-10} \cmidrule{9-10} \cmidrule{10-10} 
							&  & $N = 20$ & $N = 40$ & $N = 60$ & $N = 80$ & $N = 20$ & $N = 40$ & $N = 60$ & $N = 80$\tabularnewline
							\cmidrule{3-10} \cmidrule{4-10} \cmidrule{5-10} \cmidrule{6-10} \cmidrule{7-10} \cmidrule{8-10} \cmidrule{9-10} \cmidrule{10-10} 
							&  & \multicolumn{8}{c}{$M = 1$}\tabularnewline
							\midrule
							0.0 & 4.941913E+00\tnote{b} & 4.941906E+00 & 4.941906E+00 & 4.941906E+00 & 4.941907E+00 & 1.5E-06 & 1.5E-06 & 1.5E-06 & 1.5E-06\tabularnewline
							2.0 & 4.939727E+00 & 4.939710E+00 & 4.939711E+00 & 4.939711E+00 & 4.939712E+00 & 3.3E-06 & 3.2E-06 & 3.2E-06 & 3.2E-06\tabularnewline
							4.0 & 4.925456E+00 & 4.925396E+00 & 4.925399E+00 & 4.925400E+00 & 4.925400E+00 & 1.2E-05 & 1.1E-05 & 1.1E-05 & 1.1E-05\tabularnewline
							6.0 & 4.848762E+00 & 4.848577E+00 & 4.848594E+00 & 4.848597E+00 & 4.848599E+00 & 3.8E-05 & 3.4E-05 & 3.4E-05 & 3.3E-05\tabularnewline
							8.0 & 4.439137E+00 & 4.438838E+00 & 4.438918E+00 & 4.438935E+00 & 4.438942E+00 & 6.7E-05 & 4.9E-05 & 4.5E-05 & 4.4E-05\tabularnewline
							10.0 & 2.251772E+00 & 2.251771E+00 & 2.251771E+00 & 2.251771E+00 & 2.251772E+00 & 3.0E-07 & 3.0E-07 & 3.0E-07 & 3.0E-07\tabularnewline
							\midrule 
							&  & \multicolumn{8}{c}{$M = 2$}\tabularnewline
							\midrule
							0.0 & 4.941913E+00 & 4.941913E+00 & 4.941913E+00 & 4.941913E+00 & 4.941913E+00 & 5.8E-08 & 1.2E-08 & 4.0E-09 & 9.9E-10\tabularnewline
							2.0 & 4.939727E+00 & 4.939726E+00 & 4.939727E+00 & 4.939727E+00 & 4.939727E+00 & 1.6E-07 & 3.8E-08 & 1.4E-08 & 5.9E-09\tabularnewline
							4.0 & 4.925456E+00 & 4.925451E+00 & 4.925455E+00 & 4.925455E+00 & 4.925455E+00 & 8.6E-07 & 2.1E-07 & 9.1E-08 & 4.7E-08\tabularnewline
							6.0 & 4.848762E+00 & 4.848739E+00 & 4.848756E+00 & 4.848759E+00 & 4.848761E+00 & 4.7E-06 & 1.2E-06 & 5.5E-07 & 3.1E-07\tabularnewline
							8.0 & 4.439137E+00 & 4.439025E+00 & 4.439105E+00 & 4.439122E+00 & 4.439129E+00 & 2.5E-05 & 7.2E-06 & 3.3E-06 & 1.9E-06\tabularnewline
							10.0 & 2.251772E+00 & 2.251771E+00 & 2.251771E+00 & 2.251771E+00 & 2.251771E+00 & 3.0E-07 & 3.0E-07 & 3.0E-07 & 3.0E-07\tabularnewline
							\midrule 
							&  & \multicolumn{8}{c}{$M = 3$}\tabularnewline
							\midrule
							0.0 & 4.941913E+00 & 4.941913E+00 & 4.941913E+00 & 4.941913E+00 & 4.941913E+00 & 6.1E-08 & 1.5E-08 & 7.2E-09 & 4.2E-09\tabularnewline
							2.0 & 4.939727E+00 & 4.939726E+00 & 4.939727E+00 & 4.939727E+00 & 4.939727E+00 & 1.7E-07 & 4.3E-08 & 1.9E-08 & 1.1E-08\tabularnewline
							4.0 & 4.925456E+00 & 4.925451E+00 & 4.925455E+00 & 4.925455E+00 & 4.925455E+00 & 8.8E-07 & 2.3E-07 & 1.0E-07 & 5.7E-08\tabularnewline
							6.0 & 4.848762E+00 & 4.848739E+00 & 4.848756E+00 & 4.848759E+00 & 4.848761E+00 & 4.7E-06 & 1.2E-06 & 5.5E-07 & 3.1E-07\tabularnewline
							8.0 & 4.439137E+00 & 4.439025E+00 & 4.439105E+00 & 4.439123E+00 & 4.439129E+00 & 2.5E-05 & 7.1E-06 & 3.2E-06 & 1.8E-06\tabularnewline
							10.0 & 2.251772E+00 & 2.251771E+00 & 2.251771E+00 & 2.251771E+00 & 2.251771E+00 & 3.0E-07 & 3.0E-07 & 3.0E-07 & 3.0E-07\tabularnewline
							\bottomrule
						\end{tabular}
						\begin{tablenotes}[flushleft]
							\footnotesize
							\item[a] $x = \pm x^{\star} +10.0$. For example, if $x^{\star} = 2.0$, the results presented are valid for $x = 8.0$ and $x = 12.0$.
							\item[b] Read as 4.941913$\times10^{+00}$.
						\end{tablenotes}
					\end{threeparttable}
				}
			}
		\end{center}
		\label{p14}
	\end{table}

	\begin{table}[]
		\caption{Neutron scalar flux for Test Problem 1, with $c = 0.9$, $x_{1}=0.0\,cm$, and $x_{2} = 20.0\,cm$.}
		\begin{center}
			{\renewcommand{\arraystretch}{1.15}
				\adjustbox{angle=90}{
					\begin{threeparttable}
						\begin{tabular}{cccccc|cccc}
							\toprule 
							$x^{\star}$\tnote{a} & Solution of \cref{eq4.1} & \multicolumn{4}{c}{Nonclassical Solution (\cref{eq4.2})} & \multicolumn{4}{c}{ Relative Error}\tabularnewline
							$(cm)$ & $(neutrons/cm^{2}s)$ & \multicolumn{4}{c}{$(neutrons/cm^{2}s)$} & \multicolumn{4}{c}{}\tabularnewline
							\cmidrule{3-10} \cmidrule{4-10} \cmidrule{5-10} \cmidrule{6-10} \cmidrule{7-10} \cmidrule{8-10} \cmidrule{9-10} \cmidrule{10-10} 
							&  & $N = 20$ & $N = 40$ & $N = 60$ & $N = 80$ & $N = 20$ & $N = 40$ & $N = 60$ & $N = 80$\tabularnewline
							\cmidrule{3-10} \cmidrule{4-10} \cmidrule{5-10} \cmidrule{6-10} \cmidrule{7-10} \cmidrule{8-10} \cmidrule{9-10} \cmidrule{10-10} 
							&  & \multicolumn{8}{c}{$M = 1$}\tabularnewline
							\midrule
							0.0 & 3.238645E+01\tnote{b} & 3.238327E+01 & 3.238373E+01 & 3.238382E+01 & 3.238385E+01 & 9.8E-05 & 8.4E-05 & 8.1E-05 & 8.0E-05\tabularnewline
							2.0 & 3.192730E+01 & 3.192387E+01 & 3.192442E+01 & 3.192452E+01 & 3.192456E+01 & 1.0E-04 & 9.0E-05 & 8.6E-05 & 8.5E-05\tabularnewline
							4.0 & 3.035951E+01 & 3.035541E+01 & 3.035629E+01 & 3.035646E+01 & 3.035651E+01 & 1.3E-04 & 1.0E-04 & 1.0E-04 & 9.8E-05\tabularnewline
							6.0 & 2.703330E+01 & 2.702841E+01 & 2.702997E+01 & 2.703027E+01 & 2.703037E+01 & 1.8E-04 & 1.2E-04 & 1.1E-04 & 1.0E-04\tabularnewline
							8.0 & 2.057001E+01 & 2.056512E+01 & 2.056798E+01 & 2.056853E+01 & 2.056873E+01 & 2.3E-04 & 9.8E-05 & 7.1E-05 & 6.2E-05\tabularnewline
							10.0 & 8.290805E+00 & 8.290735E+00 & 8.290744E+00 & 8.290746E+00 & 8.290747E+00 & 8.4E-06 & 7.3E-06 & 7.1E-06 & 7.0E-06\tabularnewline
							\midrule 
							&  & \multicolumn{8}{c}{$M = 2$}\tabularnewline
							\midrule
							0.0 & 3.238645E+01 & 3.238584E+01 & 3.238630E+01 & 3.238638E+01 & 3.238641E+01 & 1.8E-05 & 4.8E-06 & 2.1E-06 & 1.2E-06\tabularnewline
							2.0 & 3.192730E+01 & 3.192655E+01 & 3.192711E+01 & 3.192721E+01 & 3.192725E+01 & 2.3E-05 & 5.9E-06 & 2.6E-06 & 1.5E-06\tabularnewline
							4.0 & 3.035951E+01 & 3.035833E+01 & 3.035921E+01 & 3.035938E+01 & 3.035943E+01 & 3.8E-05 & 9.9E-06 & 4.4E-06 & 2.5E-06\tabularnewline
							6.0 & 2.703330E+01 & 2.703119E+01 & 2.703276E+01 & 2.703305E+01 & 2.703316E+01 & 7.7E-05 & 1.9E-05 & 8.9E-06 & 5.1E-06\tabularnewline
							8.0 & 2.057001E+01 & 2.056614E+01 & 2.056901E+01 & 2.056956E+01 & 2.056976E+01 & 1.8E-04 & 4.8E-05 & 2.1E-05 & 1.2E-05\tabularnewline
							10.0 & 8.290805E+00 & 8.290790E+00 & 8.290799E+00 & 8.290801E+00 & 8.290802E+00 & 1.8E-06 & 7.4E-07 & 5.3E-07 & 4.6E-07\tabularnewline
							\midrule 
							&  & \multicolumn{8}{c}{$M = 3$}\tabularnewline
							\midrule
							0.0 & 3.238645E+01 & 3.238584E+01 & 3.238630E+01 & 3.238638E+01 & 3.238641E+01 & 1.8E-05 & 4.8E-06 & 2.1E-06 & 1.2E-06\tabularnewline
							2.0 & 3.192730E+01 & 3.192655E+01 & 3.192711E+01 & 3.192721E+01 & 3.192725E+01 & 2.3E-05 & 5.9E-06 & 2.6E-06 & 1.5E-06\tabularnewline
							4.0 & 3.035951E+01 & 3.035833E+01 & 3.035921E+01 & 3.035938E+01 & 3.035944E+01 & 3.8E-05 & 9.9E-06 & 4.4E-06 & 2.5E-06\tabularnewline
							6.0 & 2.703330E+01 & 2.703119E+01 & 2.703276E+01 & 2.703305E+01 & 2.703316E+01 & 7.7E-05 & 1.9E-05 & 8.9E-06 & 5.0E-06\tabularnewline
							8.0 & 2.057001E+01 & 2.056615E+01 & 2.056902E+01 & 2.056956E+01 & 2.056976E+01 & 1.8E-04 & 4.8E-05 & 2.1E-05 & 1.2E-05\tabularnewline
							10.0 & 8.290805E+00 & 8.290790E+00 & 8.290799E+00 & 8.290801E+00 & 8.290802E+00 & 1.8E-06 & 7.4E-07 & 5.3E-07 & 4.6E-07\tabularnewline
							\bottomrule
						\end{tabular}
						\begin{tablenotes}[flushleft]
							\footnotesize
							\item[a] $x = \pm x^{\star}+10.0$. For example, if $x^{\star} = 2.0$, the results presented are valid for $x = 8.0$ and $x = 12.0$.
							\item[b] Read as 3.238645$\times10^{+01}$.
						\end{tablenotes}
					\end{threeparttable}
				}
			}
		\end{center}
		\label{p15}
	\end{table}
	
	\begin{table}[]
		\caption{Neutron scalar flux for Test Problem 1, with $c = 0.99$, $x_{1}=0.0\,cm$, and $x_{2} = 20.0\,cm$.}
		\begin{center}
			{\renewcommand{\arraystretch}{1.15}
				\adjustbox{angle=90}{
					\begin{threeparttable}
						\begin{tabular}{cccccc|cccc}
							\toprule 
							$x^{\star}$\tnote{a} & Solution of \cref{eq4.1} & \multicolumn{4}{c}{Nonclassical Solution (\cref{eq4.2})} & \multicolumn{4}{c}{ Relative Error}\tabularnewline
							$(cm)$ & $(neutrons/cm^{2}s)$ & \multicolumn{4}{c}{$(neutrons/cm^{2}s)$} & \multicolumn{4}{c}{}\tabularnewline
							\cmidrule{3-10} \cmidrule{4-10} \cmidrule{5-10} \cmidrule{6-10} \cmidrule{7-10} \cmidrule{8-10} \cmidrule{9-10} \cmidrule{10-10} 
							&  & $N = 20$ & $N = 40$ & $N = 60$ & $N = 80$ & $N = 20$ & $N = 40$ & $N = 60$ & $N = 80$\tabularnewline
							\cmidrule{3-10} \cmidrule{4-10} \cmidrule{5-10} \cmidrule{6-10} \cmidrule{7-10} \cmidrule{8-10} \cmidrule{9-10} \cmidrule{10-10} 
							&  & \multicolumn{8}{c}{$M = 1$}\tabularnewline
							\midrule
							0.0 & 1.378376E+02\tnote{b} & 1.377670E+02 & 1.377790E+02 & 1.377813E+02 & 1.377821E+02 & 5.1E-04 & 4.2E-04 & 4.0E-04 & 4.0E-04\tabularnewline
							2.0 & 1.336506E+02 & 1.335818E+02 & 1.335940E+02 & 1.335963E+02 & 1.335972E+02 & 5.1E-04 & 4.2E-04 & 4.0E-04 & 3.9E-04\tabularnewline
							4.0 & 1.209212E+02 & 1.208579E+02 & 1.208709E+02 & 1.208733E+02 & 1.208742E+02 & 5.2E-04 & 4.1E-04 & 3.9E-04 & 3.8E-04\tabularnewline
							6.0 & 9.913735E+01 & 9.908398E+01 & 9.909822E+01 & 9.910092E+01 & 9.910188E+01 & 5.3E-04 & 3.9E-04 & 3.6E-04 & 3.5E-04\tabularnewline
							8.0 & 6.742281E+01 & 6.738474E+01 & 6.740074E+01 & 6.740378E+01 & 6.740486E+01 & 5.6E-04 & 3.2E-04 & 2.8E-04 & 2.6E-04\tabularnewline
							10.0 & 2.450188E+01 & 2.449647E+01 & 2.449737E+01 & 2.449754E+01 & 2.449760E+01 & 2.2E-04 & 1.8E-04 & 1.7E-04 & 1.7E-04\tabularnewline
							\midrule 
							&  & \multicolumn{8}{c}{$M = 2$}\tabularnewline
							\midrule
							0.0 & 1.378376E+02 & 1.378214E+02 & 1.378334E+02 & 1.378357E+02 & 1.378365E+02 & 1.1E-04 & 3.0E-05 & 1.3E-05 & 7.6E-06\tabularnewline
							2.0 & 1.336506E+02 & 1.336341E+02 & 1.336464E+02 & 1.336487E+02 & 1.336495E+02 & 1.2E-04 & 3.1E-05 & 1.4E-05 & 8.0E-06\tabularnewline
							4.0 & 1.209212E+02 & 1.209037E+02 & 1.209167E+02 & 1.209192E+02 & 1.209200E+02 & 1.4E-04 & 3.7E-05 & 1.6E-05 & 9.4E-06\tabularnewline
							6.0 & 9.913735E+01 & 9.911820E+01 & 9.913244E+01 & 9.913514E+01 & 9.913610E+01 & 1.9E-04 & 4.9E-05 & 2.2E-05 & 1.2E-05\tabularnewline
							8.0 & 6.742281E+01 & 6.740127E+01 & 6.741727E+01 & 6.742031E+01 & 6.742139E+01 & 3.1E-04 & 8.2E-05 & 3.6E-05 & 2.1E-05\tabularnewline
							10.0 & 2.450188E+01 & 2.450066E+01 & 2.450156E+01 & 2.450173E+01 & 2.450179E+01 & 4.9E-05 & 1.3E-05 & 6.1E-06 & 3.6E-06\tabularnewline
							\midrule 
							&  & \multicolumn{8}{c}{$M = 3$}\tabularnewline
							\midrule
							0.0 & 1.378376E+02 & 1.378214E+02 & 1.378335E+02 & 1.378357E+02 & 1.378365E+02 & 1.1E-04 & 2.9E-05 & 1.3E-05 & 7.6E-06\tabularnewline
							2.0 & 1.336506E+02 & 1.336341E+02 & 1.336464E+02 & 1.336487E+02 & 1.336495E+02 & 1.2E-04 & 3.1E-05 & 1.4E-05 & 8.0E-06\tabularnewline
							4.0 & 1.209212E+02 & 1.209037E+02 & 1.209167E+02 & 1.209192E+02 & 1.209200E+02 & 1.4E-04 & 3.6E-05 & 1.6E-05 & 9.4E-06\tabularnewline
							6.0 & 9.913735E+01 & 9.911821E+01 & 9.913245E+01 & 9.913515E+01 & 9.913611E+01 & 1.9E-04 & 4.9E-05 & 2.2E-05 & 1.2E-05\tabularnewline
							8.0 & 6.742281E+01 & 6.740128E+01 & 6.741728E+01 & 6.742032E+01 & 6.742140E+01 & 3.1E-04 & 8.2E-05 & 3.6E-05 & 2.0E-05\tabularnewline
							10.0 & 2.450188E+01 & 2.450066E+01 & 2.450156E+01 & 2.450173E+01 & 2.450179E+01 & 4.9E-05 & 1.3E-05 & 6.1E-06 & 3.6E-06\tabularnewline
							\bottomrule
						\end{tabular}
						\begin{tablenotes}[flushleft]
							\footnotesize
							\item[a] $x = \pm x^{\star}+10.0$. For example, if $x^{\star} = 2.0$, the results presented are valid for $x = 8.0$ and $x = 12.0$.
							\item[b] Read as 1.378376$\times10^{+02}$.
						\end{tablenotes}
					\end{threeparttable}
				}
			}
		\end{center}
		\label{p16}
	\end{table}
	
	\newpage
	\subsection{Test Problem 2}\label{s42}
	
	In the second test problem, we again consider a slab of length $X=20\,cm$, and an isotropic interior source $Q$ as defined in \cref{eq4.3}.
	However, this time we define $\sigma_t = 1.0\,cm^{-1}$, which means that neutrons now have a shorter mean free path than the one considered in Test Problem 1.
	
\begin{figure}[H]
		\centering
		\includegraphics[scale=0.67]{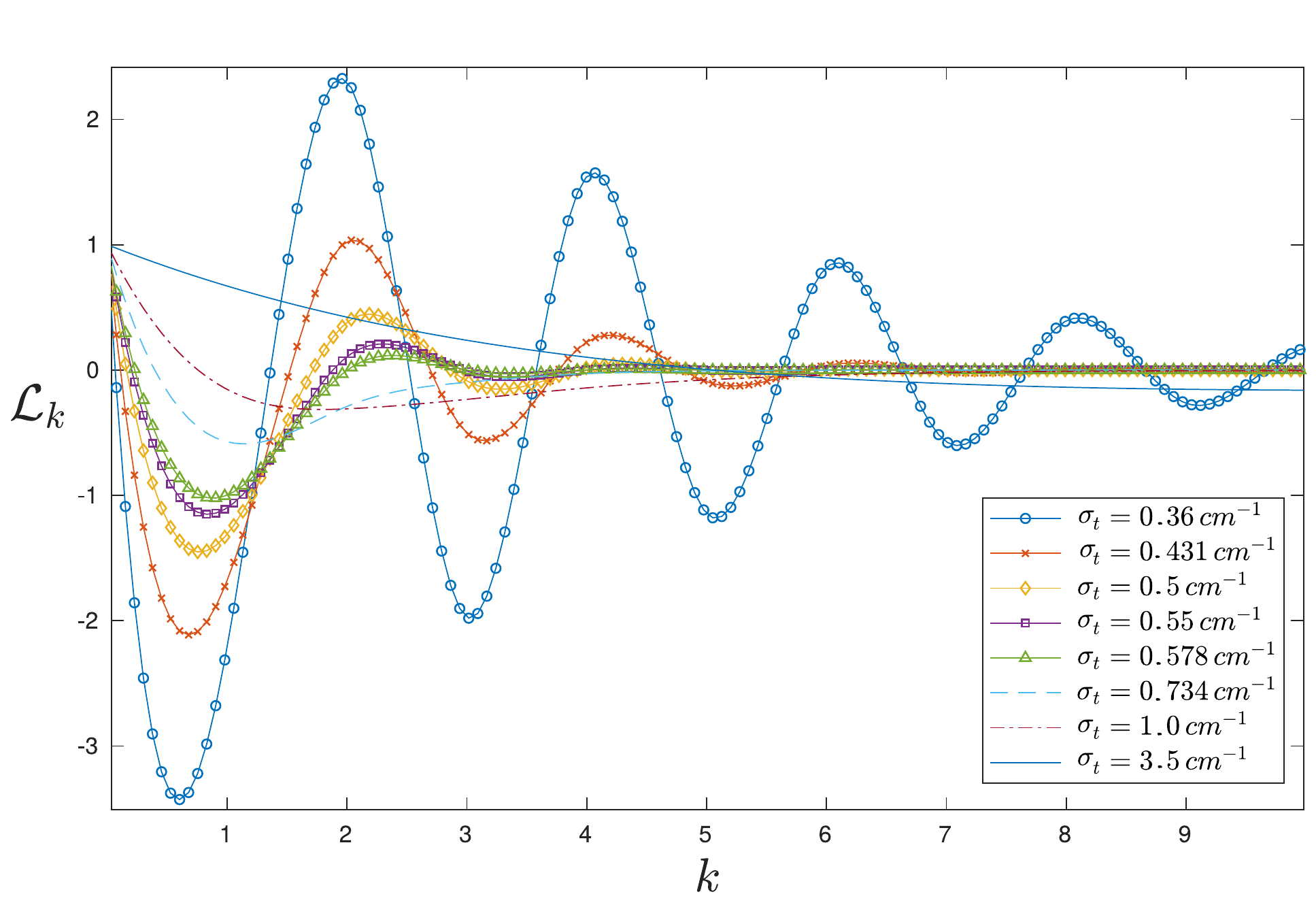}
		\caption{Profiles of the functions $\mathcal{L}_{k}$ for different values of $\sigma_{t}$.}
		\label{f2}
	\end{figure}

	Although Test Problems 1 and 2 are very similar in their choices of parameters, the numerical challenges arising in each problem are quite different.  
	As seen in \cref{s2}, the functions $\mathcal{L}_{k}$ depend on the choice of $\sigma_{t}$.
	This choice affects both the profile and the convergence rate of these functions, and consequently of the whole numerical scheme.
	Values of $\sigma_t$ that produce a sinusoidal profile and/or a low convergence rate will necessarily need a larger value for $M$ in order to generate accurate results.
	The behavior of the $\mathcal{L}_{k}$ functions for different choices of $\sigma_t$ can be seen in \cref{f2}.
	Since these functions are discrete with respect to $k$, the values depicted in \cref{f2} were interpolated to facilitate the visualization of the functions' profiles.
	
	We observe that function $\mathcal{L}_{k}$ for $\sigma_{t} = 0.578\,cm^{-1}$ converges rapidly to zero, which explains why the method generates accurate results for Test Problem 1 with a small value of $M$.
	
	In order to produce accurate solutions for problems with different values of $\sigma_t$, it is necessary to increase the values of $M$ and $N$.
	However, in the case of the ADO method, this increase produces ill-conditioned matrices for both the eigenvalue problem and the linear system of constants $\alpha$ and $\beta$ as described in \cref{s3}.
	Therefore, due to the sensitivity of the $\mathcal{L}_{k}$ functions to variations in $\sigma_t$, increasing $M$ and $N$ with the hopes of obtaining more accurate results may have the opposite effect; that is,
	the precision of the solution may be negatively affected due to the computational finite precision arithmetic.
	
	This effect is more clearly depicted in \cref{p21,p23}.
	Considering Test Problem 2 with $c=0.5$, $x_{1} = 9.5\,cm$, and $x_{2} = 10.5\,cm$, \cref{p21} presents solutions of \cref{eq4.1,eq4.2}, and the relative errors, similarly to what was done for Test Problem 1.
	On the other hand, \cref{p23} displays condition numbers that help shed more light on the overall numerical scheme.
	Namely: (i) the largest condition number obtained among the generated eigenvalues in the eigenvalue problem, which
	illustrates the sensitivity of the eigenvalues with respect to small perturbations in the matrix $\boldsymbol{BA}$ (\cref{eq3.13}); and (ii) the condition number of the linear system produced in the calculation of constants $\alpha$ and $\beta$, which gives insight into the accuracy of the constants calculated in these problems.
	
	As can be seen in \cref{p21}, increasing $M$ decreases accuracy in the solutions for all values of $N$.
	Moreover, for $M=30$, the solutions degenerate when $N$ is increased.
	This can be explained by analyzing the data in \cref{p23}: considering $M=20$ and $M=30$, we see a clear increase in the condition number of the linear systems built to calculate the constants $\alpha$ and $\beta$.
	This indicates that, from a numerical standpoint, the solutions shown in \cref{p21} do not represent the true solutions of the problem, since the condition number of the matrices is larger than the precision of the variables in which the algebraic and matrix operations are being performed (\textit{double} precision, i.e. 16 digits).
	Therefore, in order to obtain numerical results that represent a more accurate solution of this problem, it is not sufficient to choose appropriate values of $N$ and $M$.
	One also needs to ensure that the algebraic and matrix operations are performed taking into consideration an appropriate amount of precision digits.
	
	In \cref{p2u}, we present solutions of \cref{eq4.1,eq4.2}, and corresponding relative errors.
	We consider scattering ratio $c=0.5$, $M=30$, and a positive interior source $Q$ in the center of the system, with $x_1 = 9.5\,cm$ and $x_2 = 10.5\,cm$.
	These results are presented for two different precisions: 16 and 40.
	As expected, the solutions obtained with 16 digits of precision become worse as $N$ increases.
	On the other hand, the solutions generated when using 40 digits of precision maintain their accuracy when increasing $N$.

	\begin{table}[H]
		\caption{Neutron scalar flux for Test Problem 2, with $c = 0.5$, $x_{1}=9.5\,cm$, and $x_{2} = 10.5\,cm$.}
		\begin{center}
			{\renewcommand{\arraystretch}{1.15}
				\adjustbox{scale=1,angle=90}{
					\begin{threeparttable}
						\begin{tabular}{ccccc|ccc}
							\toprule 
							$x^{\star}$\tnote{a} & Solution of \cref{eq4.1} & \multicolumn{3}{c}{Nonclassical Solution (\cref{eq4.2})} & \multicolumn{3}{c}{Relative Error}\tabularnewline
							$(cm)$ & $(neutrons/cm^{2}s)$ & \multicolumn{3}{c}{$(neutrons/cm^{2}s)$} &  &  & \tabularnewline
							\cmidrule{3-8} \cmidrule{4-8} \cmidrule{5-8} \cmidrule{6-8} \cmidrule{7-8} \cmidrule{8-8} 
							&  & $N = 30$ & $N = 40$ & $N = 50$ & $N = 30$ & $N = 40$ & $N = 50$\tabularnewline
							\midrule 
							&  & \multicolumn{6}{c}{$M=20$}\tabularnewline
							\midrule
							0.0 & 1.831746E+00\tnote{b} & 1.831724E+00 & 1.831762E+00 & 1.831750E+00 & 1.2E-05 & 8.5E-06 & 2.2E-06\tabularnewline
							2.0 & 2.249508E-01 & 2.249508E-01 & 2.249508E-01 & 2.249509E-01 & 2.6E-08 & 4.2E-08 & 1.5E-07\tabularnewline
							4.0 & 1.942172E-02 & 1.942172E-02 & 1.942172E-02 & 1.942173E-02 & 2.1E-07 & 3.6E-08 & 2.4E-07\tabularnewline
							6.0 & 1.676809E-03 & 1.676809E-03 & 1.676809E-03 & 1.676809E-03 & 1.0E-07 & 2.8E-07 & 7.8E-08\tabularnewline
							8.0 & 1.445880E-04 & 1.445880E-04 & 1.445880E-04 & 1.445880E-04 & 2.9E-07 & 2.8E-07 & 5.7E-07\tabularnewline
							10.0 & 1.035481E-05 & 1.035639E-05 & 1.035569E-05 & 1.035536E-05 & 1.5E-04 & 8.4E-05 & 5.3E-05\tabularnewline
							\midrule
							&  & \multicolumn{6}{c}{$M=30$}\tabularnewline
							\midrule
							0.0 & 1.831746E+00 & 1.831613E+00 & 1.829918E+00 & 1.799102E+00 & 7.2E-05 & 9.9E-04 & 1.7E-02\tabularnewline
							2.0 & 2.249508E-01 & 2.249388E-01 & 2.248485E-01 & 2.216177E-01 & 5.3E-05 & 4.5E-04 & 1.4E-02\tabularnewline
							4.0 & 1.942172E-02 & 1.942069E-02 & 1.941289E-02 & 1.913395E-02 & 5.3E-05 & 4.5E-04 & 1.4E-02\tabularnewline
							6.0 & 1.676809E-03 & 1.676720E-03 & 1.676047E-03 & 1.651964E-03 & 5.3E-05 & 4.5E-04 & 1.4E-02\tabularnewline
							8.0 & 1.445880E-04 & 1.445802E-04 & 1.445222E-04 & 1.424456E-04 & 5.3E-05 & 4.5E-04 & 1.4E-02\tabularnewline
							10.0 & 1.035481E-05 & 1.035570E-05 & 1.035192E-05 & 1.020636E-05 & 8.5E-05 & 2.7E-04 & 1.4E-02\tabularnewline
							\bottomrule
						\end{tabular}
						\begin{tablenotes}[flushleft]
							\footnotesize
							\item[a] $x = \pm x^{\star} +10.0$. For example, if $x^{\star} = 2.0$, the results presented are valid for $x = 8.0$ and $x = 12.0$.
							\item[b] Read as 1.831746$\times10^{+00}$.
						\end{tablenotes}
					\end{threeparttable}
				}
			}
		\end{center}
		\label{p21}
	\end{table}
	
	\begin{table}[]
		\caption{Condition numbers for Test Problem 2.}\vspace*{-0.4cm}
		\begin{center}
			\begin{threeparttable}
				{\renewcommand{\arraystretch}{1.15}
					\begin{tabular}{cccc}
						\toprule 
						\multicolumn{2}{c}{Matrices} & Eigenvalue problem\tnote{a} &  Linear system\tnote{b}\tabularnewline
						\midrule
						\multirow{3}{*}{$M = 20$} & $N = 30$ & 6.699423E+15\tnote{c} & 3.108575E+26\tabularnewline
						& $N = 40$ & 4.6253121E+15 & 9.171583E+25\tabularnewline
						& $N = 50$ & 1.0877720E+16 & 1.553950E+26\tabularnewline
						\midrule 
						\multirow{3}{*}{$M = 30$} & $N = 30$ & 1.619260E+16 & 2.081604E+29\tabularnewline
						& $N = 40$ & 8.324544E+15 & 3.816123E+29\tabularnewline
						& $N = 50$ & 3.450098E+16 & 1.191783E+30\tabularnewline
						\bottomrule
				\end{tabular} }
				\begin{tablenotes}[flushleft]
					\footnotesize
					\item[a] Largest condition number among the generated eigenvalues; calculated by the \textit{condeig}($x$) function in MATLAB \cite{Matlab:2018b}.
					\item[b] Calculated by the \textit{cond}($x$) function in MATLAB \cite{Matlab:2018b}.
					\item[c] Read as 6.699423$\times10^{+15}$. 
				\end{tablenotes}
			\end{threeparttable}
		\end{center}
		\label{p23}
	\end{table}
	
	\begin{table}[]
		\caption{Neutron scalar flux for Test Problem 2, with $c = 0.5$, $M=30$, $x_{1}=9.5\,cm$, and $x_{2} = 10.5\,cm$.}
		\begin{center}
			{\renewcommand{\arraystretch}{1.15}
				\adjustbox{scale=1,angle=90}{
					\begin{threeparttable}
						\begin{tabular}{ccccc|ccc}
							\toprule 
							$x^{\star}$\tnote{a} & Solution of \cref{eq4.1} & \multicolumn{3}{c}{Nonclassical Solution (\cref{eq4.2})} & \multicolumn{3}{c}{Relative Error}\tabularnewline
							$(cm)$ & $(neutrons/cm^{2}s)$ & \multicolumn{3}{c}{$(neutrons/cm^{2}s)$} & \multicolumn{3}{c}{}\tabularnewline
							\cmidrule{3-8} \cmidrule{4-8} \cmidrule{5-8} \cmidrule{6-8} \cmidrule{7-8} \cmidrule{8-8} 
							&  & \multicolumn{6}{c}{$M = 30$}\tabularnewline
							\cmidrule{3-8} \cmidrule{4-8} \cmidrule{5-8} \cmidrule{6-8} \cmidrule{7-8} \cmidrule{8-8} 
							&  & $N = 30$ & $N = 40$ & $N = 50$ & $N = 30$ & $N = 40$ & $N = 50$\tabularnewline
							\cmidrule{3-8} \cmidrule{4-8} \cmidrule{5-8} \cmidrule{6-8} \cmidrule{7-8} \cmidrule{8-8} 
							&  & \multicolumn{6}{c}{16 significant digits of precision}\tabularnewline
							\midrule
							0.0 & 1.831746E+00\tnote{b} & 1.831613E+00 & 1.829918E+00 & 1.799102E+00 & 7.2E-05 & 9.9E-04 & 1.7E-02\tabularnewline
							2.0 & 2.249508E-01 & 2.249388E-01 & 2.248485E-01 & 2.216177E-01 & 5.3E-05 & 4.5E-04 & 1.4E-02\tabularnewline
							4.0 & 1.942172E-02 & 1.942069E-02 & 1.941289E-02 & 1.913395E-02 & 5.3E-05 & 4.5E-04 & 1.4E-02\tabularnewline
							6.0 & 1.676809E-03 & 1.676720E-03 & 1.676047E-03 & 1.651964E-03 & 5.3E-05 & 4.5E-04 & 1.4E-02\tabularnewline
							8.0 & 1.445880E-04 & 1.445802E-04 & 1.445222E-04 & 1.424456E-04 & 5.3E-05 & 4.5E-04 & 1.4E-02\tabularnewline
							10.0 & 1.035481E-05 & 1.035570E-05 & 1.035192E-05 & 1.020636E-05 & 8.5E-05 & 2.7E-04 & 1.4E-02\tabularnewline
							\midrule 
							&  & \multicolumn{6}{c}{40 significant digits of precision}\tabularnewline
							\midrule
							0.0 & 1.831746E+00 & 1.831748E+00 & 1.831762E+00 & 1.831748E+00 & 9.7E-07 & 8.6E-06 & 9.5E-07\tabularnewline
							2.0 & 2.249508E-01 & 2.249508E-01 & 2.249508E-01 & 2.249508E-01 & 4.1E-08 & 5.4E-09 & 3.4E-10\tabularnewline
							4.0 & 1.942172E-02 & 1.942172E-02 & 1.942172E-02 & 1.942172E-02 & 1.1E-07 & 2.8E-11 & 1.9E-11\tabularnewline
							6.0 & 1.676809E-03 & 1.676809E-03 & 1.676809E-03 & 1.676809E-03 & 2.0E-07 & 1.7E-09 & 1.1E-09\tabularnewline
							8.0 & 1.445880E-04 & 1.445879E-04 & 1.445879E-04 & 1.445879E-04 & 4.7E-07 & 2.2E-07 & 1.4E-07\tabularnewline
							10.0 & 1.035481E-05 & 1.035540E-05 & 1.035573E-05 & 1.035540E-05 & 5.6E-05 & 8.8E-05 & 5.6E-05\tabularnewline
							\bottomrule
						\end{tabular}
						\begin{tablenotes}[flushleft]
							\footnotesize
							\item[a] $x = \pm x^{\star}+10.0$. For example, if $x^{\star} = 2.0$, the results presented are valid for $x = 8.0$ and $x = 12.0$.
							\item[b] Read as 1.831746$\times10^{+00}$.
						\end{tablenotes}
					\end{threeparttable}
			}}
			\par\end{center}
		\label{p2u}
	\end{table}
	
	\newpage
	\section{Discussion}\label{s5}
	
	In this work we have presented a detailed study of the application of the ADO method in obtaining a numerical solution for the spectral approximation of the nonclassical transport equations.
	In this approximation the nonclassical angular flux is expanded in a series of Laguerre polynomials, resulting in a system of equations that have the same form of the classical transport equations.
	These can be solved through classical deterministic methods, whose performance needs to be analyzed for a better understanding of their behavior 
	when addressing nonclassical problems.
	We have elected in this paper to use the ADO method, which produces explicit solutions in the spatial variable.
	Moreover, the ADO method generates an eigenvalue problem whose order is half of those obtained with other conventional spectral approaches \cite{Barichello:1999:Discrete,Barichello:2011:Explicit}.
	
	In using the spectral approximation of the nonclassical transport equations, we need to deal with the $\mathcal{L}_{k}$ functions, introduced in \cref{eq1.5c}.
	These functions play an important role in the solution of the nonclassical problem.
	In \cref{s2} we have analytically calculated these functions, observing that as $k\rightarrow \infty$ they will only converge if $\sigma_{t} > \frac{\sqrt{3}}{6}$.
	This indicates that the numerical solution as generated by using the spectral approximation will diverge for the cases with $\sigma_{t} \leq \frac{\sqrt{3}}{6}$, regardless of the deterministic method used to obtain the solution.
	Therefore, modifications to the spectral approach must be explored in order to tackle problems in which the functions $\mathcal{L}_{k}$ diverge.
	This was first suggested in \cite{Vasques:2020:Spectral}, and now we have shown substantial evidence supporting that suggestion. 
	Such modifications shall be pursued in future work.
	
	It is important to point out that the nonclassical transport equation depends upon the free-path variable $s$.
	This implies that the integral that defines the functions $\mathcal{L}_{k}$ may be approximated, and evaluated only on the finite interval upon which the problem is being solved, since the particle cannot travel a distance between collisions that is larger than the domain itself.
	Thus, there is the chance that the functions $\mathcal{L}_{k}$ may diverge for a specific choice of parameters in a certain domain, and converge for the same choice of parameters when considering a smaller domain.
	However, this does not change the need to explore improvements to the spectral approach to treat diverging $\mathcal{L}_{k}$ functions.
	
	In \cref{s4} we presented numerical results for two test problems, illustrating the precision of the ADO method. In Test Problem 1, we observed that the ADO method presented high precision when solving the nonclassical problem.
	Two sets of problems were investigated, with different choices of scattering ratios and interior source intervals.
	The ADO method showed low sensitivity to changes in the scattering ratio, and its accuracy increased as $M$ and $N$ became larger.
	Moreover, when the source interval was increased, the accuracy of the ADO method also improved.
	
	In the second test problem we investigated the influence of the choice of $\sigma_{t}$ on the behavior of the 
	$\mathcal{L}_{k}$ functions, and consequently on its effect on the ADO formulation.
	We found that the truncation order $M$ of the Laguerre polynomials, needed for the generation of accurate results in the spectral approximation approach, varies with the choice of total cross section.
	This has a direct effect on the efficiency of the ADO method since the increase of $N$ and $M$ contributes to an increase in the condition number of the matrices built for this method.
	It becomes necessary, for certain cases, to use arbitrary precision libraries in order to generate numerical results that represent the true solution of the problem.
	This is showcased in \cref{p5}, in which we present the execution time and RAM allocation (\textit{Resident Set Size}) for the solutions given in \cref{p2u}.
	\begin{table}[H]
		\caption{Data on the ADO method's efficiency when applied to Test Problem 2.} \vspace*{-0.5cm}
		\begin{center}
			{\renewcommand{\arraystretch}{1.15}
				\adjustbox{scale=0.95}{
					\begin{threeparttable}
						\begin{tabular}{ccccc}
							\toprule 
							Significant digits of precision & \multicolumn{2}{c}{Experiments} & Time of execution (sec) & RAM allocation\tnote{a}  (kB)\tabularnewline
							\midrule
							\multirow{3}{*}{16} & \multirow{6}{*}{$M = 30$} & $N = 30$ & 7.388625E+00\tnote{b} & 2.622930E+05\tabularnewline
							&  & $N = 40$ & 1.770191E+01 & 4.542840E+05\tabularnewline
							&  & $N = 50$ & 3.807081E+01 & 5.193640E+05\tabularnewline
							\cmidrule{1-1} \cmidrule{4-5} \cmidrule{5-5} 
							\multirow{3}{*}{40} &  & $N = 30$ & 4.148346E+03 & 9.480600E+05\tabularnewline
							&  & $N = 40$ & 9.944762E+03 & 1.679036E+06\tabularnewline
							&  & $N = 50$ & 1.925917E+04 & 2.146800E+06\tabularnewline
							\bottomrule
						\end{tabular}
						\begin{tablenotes}[flushleft]
							\footnotesize
							\item[a] \textit{Resident Set Size}.
							\item[b] Read as 7.388625$\times10^{+00}$.
							\item[c] All calculations were performed on a notebook with the following configuration: Intel(R) Core(TM) i5-5200U CPU@ 2.20GHz, 8GB RAM.
						\end{tablenotes}
					\end{threeparttable}
			}}
			\par\end{center}
		\label{p5}
	\end{table}
	
	As future work, we intend to study approaches that yield a smaller condition number of the matrices built to solve these problems, such as to explore a potential hybrid algorithm that combines the ADO method with the Response Matrix method \cite{Silva:2020:Response,Moraes:2020:Estimation}.
	In addition, we aim to explore different representations of the nonclassical angular flux currently given by \cref{eq1.4a}, in order to attain modified forms of the $\mathcal{L}_{k}$ functions with a better convergence rate.
	This would prevent, in some cases, the need to work with high values of $M$, in a similar fashion to what we have seen in Test problem 1.
	In other words, $\mathcal{L}_{k}$ functions with a faster convergence rate will prevent the need to use high precision algebraic and matrix calculations, improving the overall efficiency of the computer code.
	
	\section*{Acknowledgments}
	
	This study was financed in part by the Coordena\c c\~ao de Aperfei\c coamento de Pessoal de N\'ivel Superior - Brasil (CAPES) - Finance Code 001, and Funda\c c\~ao Carlos Chagas Filho de Amparo \`a Pesquisa do Estado do Rio de Janeiro - Brasil (FAPERJ).
	L.R.C.~Moraes, L.B.~Barichello and R.C.~Barros acknowledge support from Conselho Nacional de Desenvolvimento Cient\'ifico e Tecnol\'ogico - Brasil (CNPq). L.R.C.~Moraes also would like to express his gratitude to the Graduate Program in Applied Mathematics of Universidade Federal do Rio Grande do Sul - Brasil, due to the support and care provided during the development of this work.
	R.~Vasques acknowledges support under award number NRC-HQ-84-15-G-0024 from the Nuclear Regulatory Commission.
	
	
	\bibliography{Notas}
	
\end{document}